\documentclass[%
 reprint,
 superscriptaddress,
 amsmath,amssymb,
 aps,
floatfix,
]{revtex4-2}
\usepackage[english]{babel}
\usepackage{graphicx}
\usepackage{dcolumn}
\usepackage{bm}
\usepackage{comment}
\usepackage{xcolor}
\usepackage{amsmath}
\usepackage{braket}
\usepackage[normalem]{ulem}
\usepackage{float}
\usepackage[colorlinks,citecolor=red,urlcolor=blue,bookmarks=false,hypertexnames=true]{hyperref} 

\usepackage[autostyle]{csquotes}
\MakeOuterQuote{"}

\usepackage{todonotes}

\begin{document}

\title{All-Photonic Artificial Neural Network Processor Via Non-linear Optics}

\author{Jasvith Raj Basani}%
\affiliation{Department of Electrical and Electronics Engineering, Birla Institute of Technology and Science, Pilani, Hyderabad Campus, Telangana, 500078, India}%
\affiliation{Department of Electrical and Computer Engineering, Institute for Research in Electronics and Applied Physics, and Joint Quantum Institute, University of Maryland, College Park, MD 20742, USA}

\author{Mikkel Heuck}%
\affiliation{ Department of Electrical Engineering and Computer Science, Massachusetts Institute of Technology, 77 Massachusetts Avenue, Cambridge, Massachusetts 02139, USA}%
\affiliation{Department of Electrical and Photonics Engineering, Technical University of Denmark, Building 343, 2800 Kgs. Lyngby, Denmark}%

\author{Dirk R. Englund}%
\affiliation{ Department of Electrical Engineering and Computer Science, Massachusetts Institute of Technology,
77 Massachusetts Avenue, Cambridge, Massachusetts 02139, USA}%

\author{Stefan Krastanov}%
\affiliation{ Department of Electrical Engineering and Computer Science, Massachusetts Institute of Technology,
77 Massachusetts Avenue, Cambridge, Massachusetts 02139, USA}%

\date{\today}

\begin{abstract}
Optics and photonics has recently captured interest as a platform to accelerate linear matrix processing, that has been deemed as a bottleneck in traditional digital electronic architectures. In this paper, we propose an all-photonic artificial neural network processor wherein information is encoded in the amplitudes of frequency modes that act as neurons. The weights among connected layers are encoded in the amplitude of controlled frequency modes that act as pumps. Interaction among these modes for information processing is enabled by non-linear optical processes. Both the matrix multiplication and element-wise activation functions are performed through coherent processes, enabling the direct representation of negative and complex numbers without the use of detectors or digital electronics. Via numerical simulations, we show that our design achieves a performance commensurate with present-day state-of-the-art computational networks on image-classification benchmarks. Our architecture is unique in providing a completely unitary, reversible mode of computation. Additionally, the computational speed increases with the power of the pumps to arbitrarily high rates, as long as the circuitry can sustain the higher optical power. 
\end{abstract}

\maketitle


\section{\label{sec:level1}Introduction}

\begin{figure*}
    \centering
    \includegraphics[width=\textwidth]{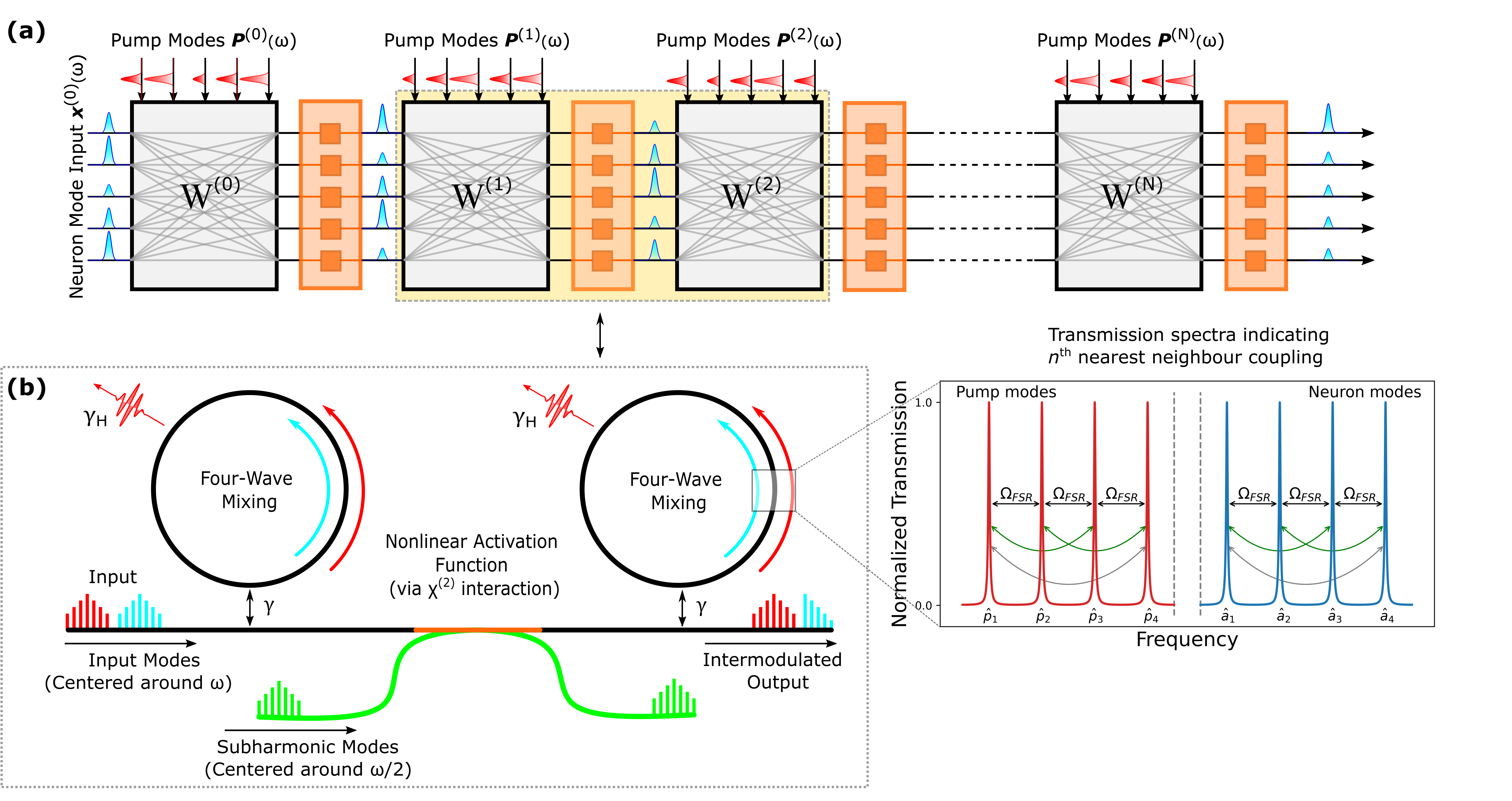}
    \caption{\textbf{(a)} Schematic of the neural network represented as a sequence of $N$ layers. The information being processed is encoded in the amplitudes of frequency states of neuron modes (blue), while the linear transformations $W^{(i)}$ is implemented via strong classical pump modes (red). The nonlinear activation function (given by orange blocks) occurs during propagation between layers via nonlinear optical interactions of the neuron modes with external subharmonic pump modes (green). \textbf{(b)} In this inset, two consecutive layers with a nonlinearity are shown to be made of microring resonators connected via a waveguide. Each ring resonator is coupled to the waveguide with a coupling constant $\gamma$ and experiences internal losses $\gamma_\mathrm{H}$. The transmission spectra of the microring resonator is shown alongside, where $n^{\mathrm{th}}$ nearest neighbour pump and neuron modes are coupled (given by green and grey arrows).} 
    \label{ring_transmission}
\end{figure*}

The last decade has witnessed phenomenal advances in the domain of machine learning, with applications ranging from natural language processing~\cite{NLP}, structural biology~\cite{protein} and even game playing~\cite{muzero}. With the growing accessibility of large datasets and larger computational power, machine learning models have been increasing in complexity to tackle a multitude of problems. The requirement for better performance in these networks has necessitated the development of hardware accelerators, specifically for the training of deep neural networks. Recently, tailored digital electronic architectures, such as Graphic Processing Units (GPUs)~\cite{GPU} and application-specific integrated circuits such as Google's Tensor Processing Units~\cite{TPU}, IBM TrueNorth~\cite{IBM}, and Intel Nervana~\cite{Intel} have been introduced to accelerate the training and inference of machine learning models. These devices still do, however, require enormous energy resources and can be uneconomical at tackling problems with large computational complexity. Recently, with advances in silicon photonics~\cite{SiP, SiP_2}, optical computing has been introduced as an attractive platform to carry out large-scale computational schemes. Properties of light, such as coherence and superposition, blended with the vast array of CMOS-compatible optical devices has made photonics a fruitful direction of exploration for efficiently and effectively implementing computational schemes.

Photonic implementations of neural networks have been proposed and realized both in free-space environments and via photonic integrated circuitry, with breakthroughs in spiking neural networks~\cite{spike} and photonic reservoir processing~\cite{reservoir}. The photonic platform has garnered interest from scientists and engineers alike, to leverage the massive parallelism being offered by the multiple degrees of freedom of light (wavelength, polarization, phase etc.). A problem faced by even the most optimized electronic architectures is the expenditure of energy for data movement as opposed to logical operations~\cite{expend_1, expend_2}. Photonic solutions, on the other hand, greatly reduce energy consumption, both in terms of data-transfer and computational operations by performing linear (and some non-linear) transformations via passive optical interaction. Moreover, linear matrix transformations have been recorded at rates exceeding 100 GHz~\cite{speed}. Advances in nanophotonics have allowed us to implement bulk optical nonlinearities readily, with minimal latency. 

Remarkably, the construction of such neural networks comprises of only two fundamental components -- linear matrix multiplication to serve as an interconnect between consecutive layers, followed by a non-linear (sigmoid) activation function. In this paper, we propose a new architecture for a fully photonic implementation of artificial neural networks based on non-linear optical intermodulation. In contrast to previous approaches~\cite{ONN_histories_1, ONN_histories_2, ONN_histories_3, ONN_histories_4, ONN_histories_5}, we encode information in the complex amplitudes of frequency states that act as neurons, in a multi-mode cavity. Furthermore, information regarding the linear transformations that the neuron modes undergo is encoded in the amplitudes of controlled pump modes. General matrix-vector and matrix-matrix multiplications are enabled via the process of Four-Wave Mixing (FWM)~\cite{FWM}, which has been used extensively for intermodulation among multiple frequency modes \cite{FWM_app}. This approach lets us represent negative (or even complex) activation values, a problem plaguing other optical approaches.
 
Unlike other optical~\cite{NL_O_1, NL_O_2, NL_O_3, NL_O_4} and opto-electronic~\cite{NL_OEO_1, NL_OEO_2, NL_OEO_3, NL_OEO_4, NL_OEO_5} approaches, we also propose a scheme to perform the elementwise sigmoid activation function coherently via a nonlinear optical process. Our scheme can represent activation functions acting on negative and even complex numbers, without passing through a detector and electronic digital computer.
 
Our design can be made rapidly re-programmable as well. The protocol we propose can be realized on microring resonators, allowing easy fabrication, via well-established lithography techniques. Moreover, the entirety of the computation performed by the proposed hardware is, in principle, reversible and unitary, opening up many possibilities for low-power (even reversible) computation and training. Lastly, the rate at which the matrix-multiplication operations are performed scales with the pump power, hence providing for extremely fast operations, as long as the circuitry can tolerate high-power control pulses.

This paper is organized as follows -- in the following Section~\ref{sec:matrix}, we introduce two schemes for matrix multiplication, the methods of \textit{passive} and \textit{active} coupling of "neuron" pulses in a multimode optical cavity. We discuss the Hamiltonian and matrix-transformation implemented by the optical cavity and establish the time-dynamics of the neuron modes. The limitations of the available operations and methods for overcoming these limitations are discussed and benchmarked on the Iris linearly-separable dataset~\cite{Iris}. Section~\ref{sec:activation} discusses our implementation of the non-linear activation function. In Section~\ref{sec:casestudy}, we perform simulations to train our neural network accelerator on the MNIST dataset~\cite{MNIST} to illustrate the performance of our hardware design in different parameter regimes. Our paper concludes with a discussion of the results, broader impact and scope for the extension and experimental realization of this work.

\section{\label{sec:matrix}Programmable Transformations via Four Wave-Mixing}

Deep neural networks (DNNs) are a class of artificial neural networks that, fundamentally, consist of multiple stacked layers of neurons, each connected via a matrix multiplication ($\vec{x} \mapsto W\vec{x}$) and an element-wise non-linear activation function ($x_{i} \mapsto \sigma (x_{i})$). For a DNN of arbitrary depth, the input to the $(k + 1)^{\mathrm{th}}$ layer is related to the input of the $k^{\mathrm{th}}$ layer as: 

\begin{equation}
    x_{i}^{(k + 1)} = \sigma \left( \sum_{j} W_{i,j}^{(k)}x_{j}^{(k)} \right)
\end{equation}

We propose realizing the matrix multiplication by $W^{(k)}$ in a multi-mode optical cavity. For instance, consider an optical cavity implemented as a microring resonator that supports a frequency comb in the telecommunication range (around 1550 nm). The frequency states supported by the microring resonator are chosen to be either "pump" or "neuron" modes, that interact with each other via the process of Four-Wave Mixing (FWM). Our design encodes information to be processed in the complex amplitudes of the neuron modes, while the matrix-multiplication operations are enabled by interaction with controlled pump modes. With FWM being an inherently third order non-linear optical process, the microring resonator will have to be fabricated from a material that facilitates the third order nonlinear optical response described with a large $\chi^{(3)}$ susceptibility coefficient. In the context of a traditional, fully-connected neural network, the weights that act as interconnects between layers are encoded in the strength of the pumps.

\subsection{Method of Passive Coupling}

Our first protocol for implementing a fully-connected neural net layer will employ \textit{propagating} neuron modes flying past a microring resonator as depicted in Fig.~\ref{ring_transmission}. Their mixing will be enabled by a FWM interaction with the control pumps in the resonator. To introduce the mechanism, consider first a resonator with only four modes, i.e. two neural modes and two pump modes. The lower two modes are the pumps that drive the system, denoted by operators ($\hat{p}_{1}, \hat{p}_{2}$). The two higher-frequency modes act as neurons, denoted by ($\hat{a}_{1}, \hat{a}_{2}$). The Hamiltonian associated with the interaction of the four waves is: 

\begin{equation}
    \hat{H} = \hbar \chi \left( \hat{p}_{1}\hat{p}_{2}^{\dagger}\hat{a}_{1}\hat{a}^{\dagger}_{2} \right) + \mathrm{H.c.}
    \label{hamiltonian_FWM}
\end{equation}

The coupling coefficient $\chi$ determines the strength of interaction, incorporating effects from several parameters including the nonlinear susceptibility of the material of our cavity, phase matching, and mode volume realized in the cavity. The pumps are assumed to be strong classical modes of light and their operators can be replaced by a classical complex amplitude $\hat{p}_i \mapsto p_{i}=\sqrt{\langle \hat{n}_i \rangle}e^{i\theta}$, involving the expectation value of the number of photons $n_i$ in the given pump mode and its phase $\theta$. Furthermore, these pumps are much stronger than the other modes and hence are non-depletive. We assume that the resonances of the modes obey the FWM energy matching condition, such that $\omega_{p_2} - \omega_{p_1} = \omega_{a_2} - \omega_{a_1}$. The neuron and pump modes couple from the waveguide into the microring resonator at fixed coupling rate of $\gamma$, for simplicity taken to be the same for all modes. The total loss rate is $\Gamma = \gamma + \gamma_\mathrm{H}$, where $\gamma_\mathrm{H}$ is the internal (intrinsic) loss rate. The time-dynamics of the modes can be solved using Coupled Mode Theory. The coupled amplitude equations for this system \cite{FWM_equations, CMT} are:
\begin{equation}
    \dot{P}_{i} = 0 = -\frac{\Gamma}{2}P_{i} - \sqrt{\gamma}S_{\textrm{in}, P_{i}}
    \label{eq:pump_powers}
\end{equation}
\begin{multline}
    \frac{\mathrm{d}A_{1}}{\mathrm{d}t} = \left( - \frac{\Gamma}{2} + i\chi|P_{1}|^{2} + i\chi|P_{2}|^{2} \right) A_{1} \\ + \left( \chi P_{1}P_{2}^{*} \right) A_{2} - \sqrt{\gamma}S_{\mathrm{in}, 1}
\end{multline}
\begin{multline}
    \frac{\mathrm{d}A_{2}}{\mathrm{d}t} = \left( - \frac{\Gamma}{2} + i\chi|P_{1}|^{2} + i\chi|P_{2}|^{2} \right)A_{2} \\ - \left( \chi P_{1}^{*}P_{2} \right) A_{1} - \sqrt{\gamma}S_{\mathrm{in}, 2}
\end{multline}
\begin{equation}
    S_{\textrm{out}, i} = S_{\textrm{in}, i} + \sqrt{\gamma}A_{i},
    \label{sout}
\end{equation}
where $A_{i}$ and $P_{i}$ represent, respectively, the amplitude of the $i^{\mathrm{th}}$ neuron mode and the amplitude of the $i^{\mathrm{th}}$ pump mode \emph{inside of the resonator}. The pump amplitudes are set to a scale much higher than the scale of the neuron activations, to permit neglecting the direct neuron-neuron interactions. The encoded data is introduced into the system via the input waveguide mode, denoted by $S_{\mathrm{in}, i}$ (representing the activation values of the neurons).
The output neuron modes, after interacting in the ring, are denoted by $S_{\mathrm{out}, i}$. The $P_{i}$ values might need a correction to account for non-linear interactions purely between the pumps, however, this is a straightforward matrix inversion problem that does not affect the dynamics discussed below.

Extending this formalism to $N$ neurons comprising a single layer of a neural network, we see that pumps which are $n^{\mathrm{th}}$ nearest neighbours (i.e. have a frequency difference of $n \times \Omega_{\mathrm{FSR}}$ for a ring with the free spectral range $\Omega_{\mathrm{FSR}}$) couples all the neuron modes at that frequency difference. This gives rise to cross coupling terms, and hence the modified coupled amplitude equations for the $i^{\mathrm{th}}$ neuron mode can thus be written as:

\begin{multline}
    \frac{\mathrm{d} A_{i}}{\mathrm{d}t} = \left( -\frac{\Gamma}{2} + i\chi \sum_{m = 1}^{N}|P_{m}|^{2} \right)A_{i} \\ - \chi \left[ \sum_{j>i}^{N} \sum_{k = 1}^{j - 1} \left( P_{k}P_{k + j - i}^{*} \right) A_{j}  - \sum_{j<i}^{i - 1} \sum_{k = 1}^{j} \left( P_{k}^{*}P_{k + i - j} \right) A_{j} \right] \\ - \sqrt{\gamma}S_{\mathrm{in}, i}
    \label{cc_eq}
\end{multline}
Without loss of generality, we make the assumption that the first pump $P_{1}$ is much stronger than the other pumps, allowing for easier experimental calibration and permitting us to neglect the cross coupling terms, leading to:

\begin{multline}
    \frac{\mathrm{d} A_{i}}{\mathrm{d}t} = \left( -\frac{\Gamma}{2} + i\chi|P_{1}|^{2} \right) A_{i} - \chi \left[ \sum_{j>i}^{N} \left( P_{1}P_{j}^{*} \right) A_{j}  \right. \\ \left. - \sum_{j<i}^{i - 1} \left( P_{1}^{*}P_{j} \right) A_{j} \right] - \sqrt{\gamma}S_{\mathrm{in}, i}
\end{multline}
which together with Eq.~\eqref{sout} lets us rewrite the system of coupled mode equations in a matrix form:

\begin{equation}
    \vec{S}_{\mathrm{out}} = \vec{S}_{\mathrm{in}} + \sqrt{\gamma} \left[ \textbf{P}^{-1} \left( \dot{\vec{A}} + \sqrt{\gamma}\vec{S}_{\mathrm{in}} \right) \right],
    \label{eq:sout_adot}
\end{equation}
where the matrix $\textbf{P}$ has constant diagonals (also known as a Toeplitz matrix). $\textbf{P}$'s $n^{\mathrm{th}}$ off-diagonal would have the value $P_{1}P_n$. From this model, we see that the amplitudes of the output modes depend on the inverse of the matrix $\textbf{P}$, i.e., on the pump amplitudes that encode the linear operation being performed.

\begin{figure}[t]
    \centering
    \includegraphics[width=\columnwidth]{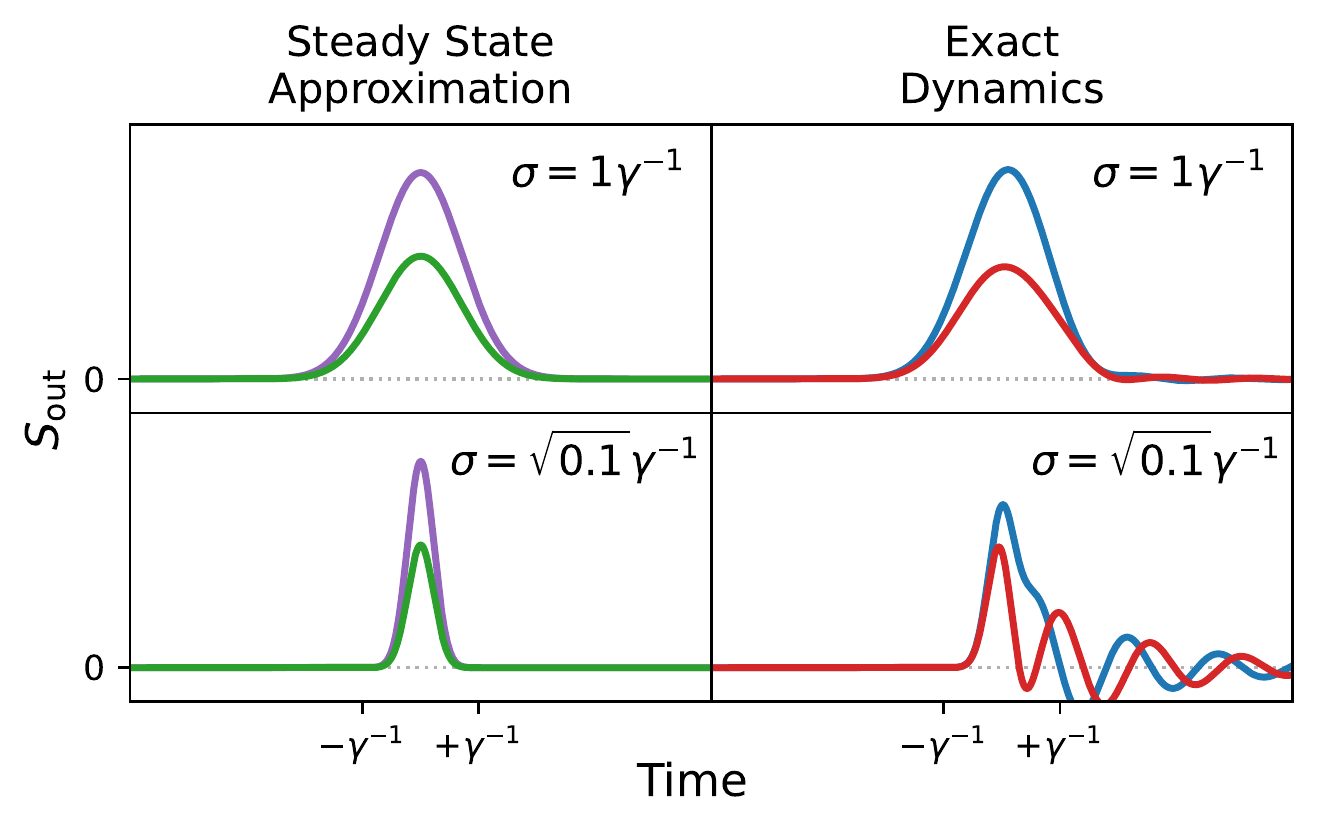}
    \caption{Comparison of the steady-state model and the full model for pulse with Gaussian envelope of different durations. The first column illustrates the correct profile of $S_{\mathrm{out}}$, while the second column show the profile as predicted by a steady-state model. For pulses much shorter than $1/\gamma$ we see the breakdown of the steady-state model.} 
    \label{fig:D_vs_SS}
\end{figure}

A deep neural network would typically consist of several layers, which in our case would be implemented by cascading multiple microring resonators consecutively. To enable repeated application of such a transformation, we need to ensure that the temporal envelope of the pulse does not vary significantly as it undergoes transformations through FWM. Assuming the $S_{\mathrm{in}}$ pulses to have a Gaussian temporal envelope, we can preserve the Gaussian shape of the output pulses $S_{out}$ if the pulses are much longer than $1/\gamma$. For pulses with a large enough duration, we can make the adiabatic elimination $\dot{\vec{A}} = 0$, allowing us to work in the steady-state regime. We illustrate this approximation in Fig.~\ref{fig:D_vs_SS} by comparing the solution of the steady-state model with the solution of the full dynamics. As the length of the input pulses increases the steady-state model begins to closely resemble the model of the full dynamics. This approximation allows us to simplify Eq.~\eqref{eq:sout_adot} into $\vec{S}_{\mathrm{out}} = \left( \mathbb{I}_{N} + \gamma \textbf{P}^{-1} \right)\vec{S}_{\mathrm{in}} = \textbf{T}\vec{S}_{\mathrm{in}}$, where $\textbf{T}$ is given in Eq.~\eqref{toeplitz}. $\mathbb{I}_{N}$ is the $N$-dimensional identity matrix.

The Toeplitz nature of the $N \times N$ matrix $\textbf{P}$ gives us only $N$ degrees of freedom, as opposed to $N^{2}$ degrees of freedom encoded in the weights of a fully-connected deep neural network. This implies that the transformation via a single layer of the form $\textbf{T}$, would span only a fraction of the space that would otherwise be spanned by the full group of unitary transformations. To quantify the group of operations that can be spanned by matrices of the form $\textbf{T}$, we introduce the concept of expressivity.

\begin{widetext}
\begin{equation}
\begin{aligned}
     \textbf{T} & = \mathbb{I}_{N} + \gamma \textbf{P}^{-1}\\
     & = \mathbb{I}_{N} + 
    \gamma 
\begin{pmatrix}
\begin{array}{ccccccccc}
    -\Gamma/2 + i\chi|P_{1}|^{2} & P_{1}P_{2}^{*}\chi & P_{1}P_{3}^{*}\chi & & & \hdots & P_{1}P_{N}^{*}\chi \\
    -P_{1}^{*}P_{2}\chi & -\Gamma/2 + i\chi|P_{1}|^{2} & P_{1}P_{2}^{*}\chi &  & & \hdots & P_{1}P_{N-1}^{*}\chi \\
    -P_{1}^{*}P_{3}\chi & -P_{1}^{*}P_{2}\chi &  -\Gamma/2 + i\chi|P_{1}|^{2} & \ddots &  & \hdots &  P_{1}P_{N-2}^{*}\chi \\
    \vdots & \ddots &  \ddots &  \ddots & & & \vdots\\
    \vdots & & \ddots &  \ddots &  & & \vdots \\
    & & & & & & \\
    -P_{1}^{*}P_{N-1}\chi & \hdots & & -P_{1}^{*}P_{2}\chi &  &  -\Gamma/2 + i\chi|P_{1}|^{2} & P_{1}P_{2}^{*}\chi\\
    -P_{1}^{*}P_{N}\chi & \hdots & & -P_{1}^{*}P_{3}\chi &  &  -P_{1}^{*}P_{2}\chi & -\Gamma/2 + i\chi|P_{1}|^{2}
\end{array}
\end{pmatrix}^{-1}
\label{toeplitz}
\end{aligned}
\end{equation}
\end{widetext}

\begin{figure*}[htp!]
\centering
\includegraphics[width = \textwidth]{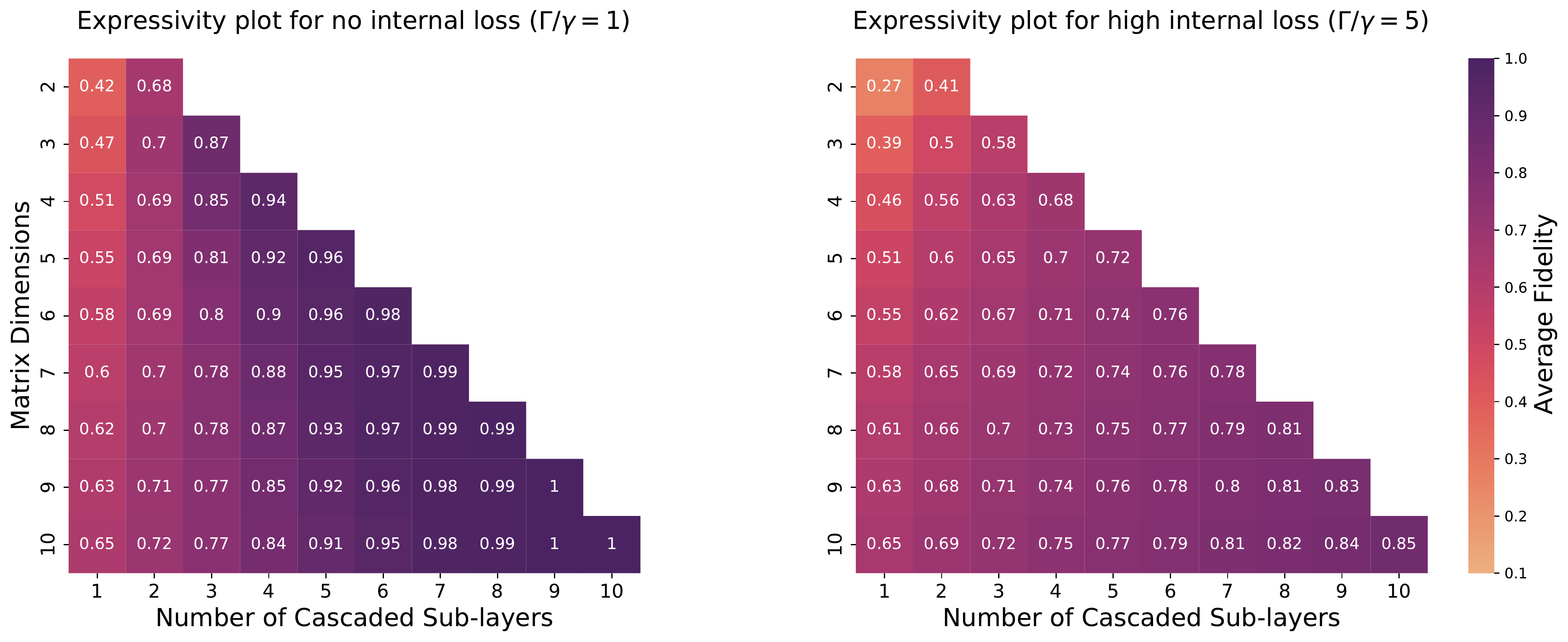}
\caption{\label{Passive_Express}Quantifying the expressivity of the transformation $\prod \textbf{T}$ in different parameter regimes of the \emph{passive coupling scheme}. Each plot displays the average expressivity as we vary the number of sub-layers (the horizontal axis) for a given matrix dimension (the vertical axis).  On the left, the expressivity at no internal loss ($\Gamma/\gamma = 1$) reaches unity at sufficiently many sub-layers. On the right, the expressivity at high loss ($\Gamma/\gamma = 5$) is consistently lower.}
\end{figure*}

The expressivity is the average fidelity with which a parametrized $\textbf{T}$ can represent an arbitrary unitary operation $\textbf{U}$. Numerically, we estimate the expressivity by sampling $M$ Haar-random unitaries $\{\textbf{U}_i\}_{1\le i\le M}$ and for each one we use gradient descent to find the $\textbf{T}_{i}$ which approximates it most closely. We estimate the expressivity as
\begin{equation}
    \mathbb{F} = 1 - \frac{1}{M}\sum_{i=1}^M\sqrt{\textrm{tr} \left[\left( \textbf{T}_{i} - \textbf{U}_{i} \right)\left( \textbf{T}_{i} - \textbf{U}_{i} \right)^{\dagger} \right] },
    \label{eq:fid}
\end{equation}
which both accounts for imperfections due to losses (deviations from unitarity) and insufficient degrees of freedom.

The transformation performed by a single layer, i.e., a single matrix of the form $\textbf{T}$ does not reach expressivity large enough to perform arbitrary unitary transformations. Here, we introduce the concept of sub-layers, wherein a single layer would consist of several non-commuting cascaded matrices of the form $\textbf{T}$ so as to produce a single compound transformation. Physically, this would require multiple subsequent ring resonators, one per sub-layer. By introducing multiple sub-layers into a  layer, i.e., multiplying multiple matrices in the form of $\textbf{T}$, we can span larger groups of operations. By estimating the expressivity of these compound operations as a function of matrix dimension and number of  sub-layers, we see that for larger matrices, at higher sub-layers, the expressivity reaches unity as illustrated in Fig.~\ref{Passive_Express}. This implies that by cascading multiple matrices in a single layer, we can span the group of unitary operations.

A factor that negatively influences the expressivity is the presence of loss, $\gamma_{\mathrm{H}}$. Up to this point, we have neglected the internal losses, i.e., $\gamma_{\mathrm{H}}=0$, thus working in the parameter regime $\Gamma / \gamma = 1$. As we see in Eq.~\eqref{toeplitz}, the diagonal of $\textbf{P}$ contains the total loss rate of each neuron mode. To illustrate the influence of intrinsic losses $\gamma_\mathrm{H}$ on the expressivity, we perform the same estimation in the parameter regime where $\gamma_\mathrm{H} > 0$. The results of the optimization indicates that even at a higher number of sub-layers, the expressivity does not reach unity. Due to loss, the compound operation does not span the group of unitary operations.

\begin{figure}
\centering
\includegraphics[width = \columnwidth]{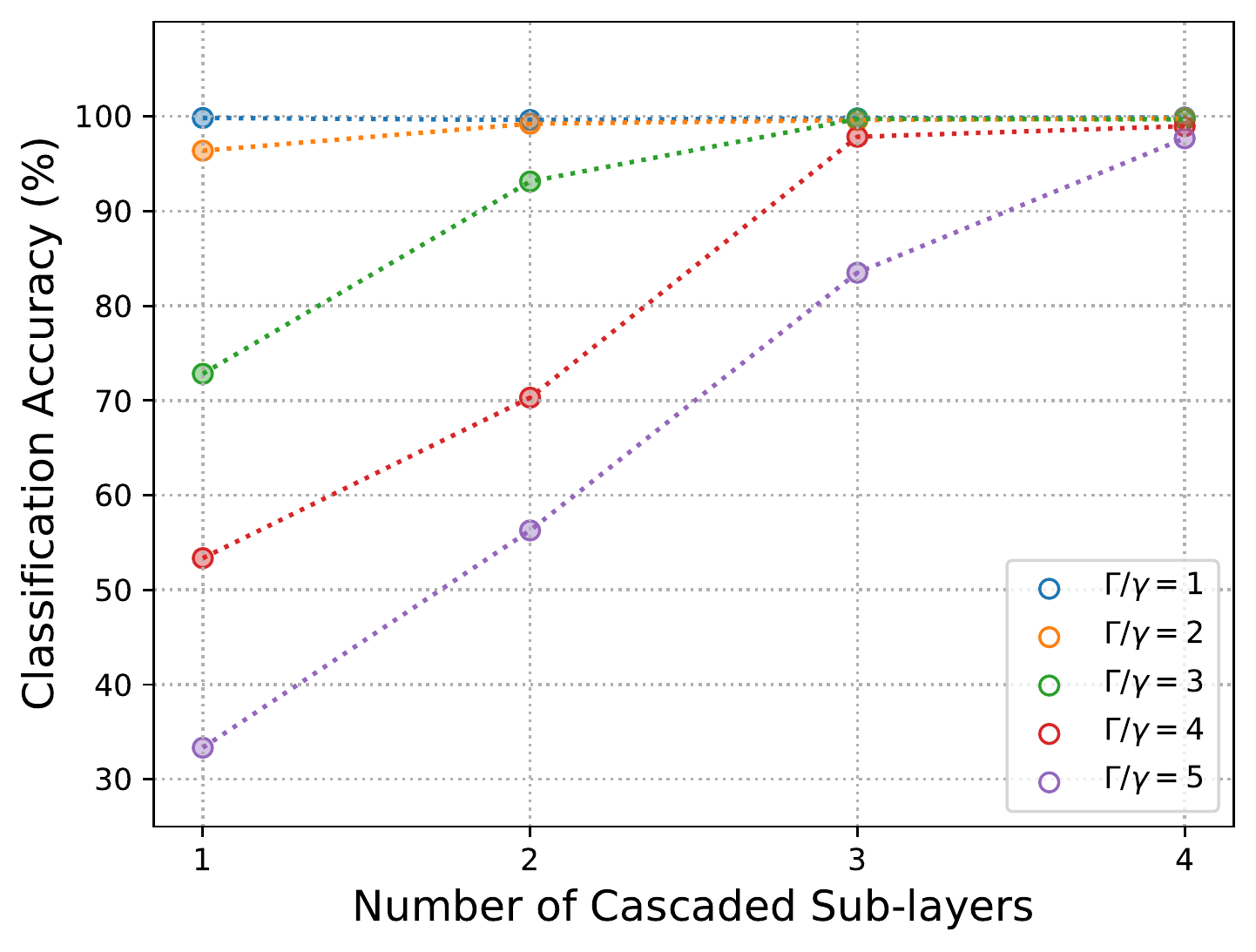}
\caption{\label{fig:Iris_passive}The classification accuracy on the Iris dataset as we vary the number of cascaded sub-layers for different $\Gamma/\gamma$ ratios for the \emph{passive coupling scheme}. A larger number of sub-layers improves the classification accuracy, while a larger $\Gamma/\gamma$ ratio is detrimental to the performance of the network.}
\end{figure}

We illustrate the holistic effect of varying the number of sub-layers and the loss ratio $\Gamma / \gamma$ through the machine learning task of linear classification of the benchmark Iris dataset in Fig.~\ref{fig:Iris_passive}. This dataset consists of only 4 features and 3 output classes, with one of the classes being linearly separable from the other two. The architecture of our neural network is a single compound layer without an elementwise sigmoid activation, between the input features and the predicted classes. We vary the number of sub-layers in the compound layer and find that for all $\Gamma / \gamma$ ratios, a larger expressivity (more sub-layers) indicates a better performance in the classification. Moreover, for a given number of sub-layers, higher losses are detrimental to the performance of the network.

Transformations of the form $\textbf{T}$ can be realized via three-wave mixing~\cite{TWM} as well, with a single pump mode instead of two as proposed above. Solving for the transformation matrix $\textbf{T}$ gives us a similar result to the one presented above (three-wave mixing does not give rise to cross-coupling between different neuron modes). The Hamiltonian associated with the interaction of the three interacting waves would be $\hat{H} = \chi \left( \hat{p} \hat{a} \hat{b}^{\dagger} \right) + \mathrm{H.c.}$, where $\hat{p}$ is the single pump mode. These modes obey the energy matching condition that $\omega_{p} = \omega_{b} - \omega_{a}$. Experimentally implementing this system, however, presents engineering challenges in the design of the microring resonator. The energy matching condition requires the frequency of the pump mode to be equal to the difference in frequencies of the neuron modes. This would result in pump modes operating at frequencies much smaller than the neuron modes, i.e., integer multiples of the Free Spectral Range (FSR) of the microring resonator. This ring would have to support modes over multiple octaves. Spanning across multiple octaves gives rise to differences in refractive indices and Q-factors for modes at different frequencies. This leads to difficulties in maintaining the resonance condition and phase matching required for high-efficiency wave-mixing. Alternatively, pump and neuron modes across multiple octaves could be implemented as an electro-optic frequency comb~\cite{EO_comb}; this approach would, however, be limited by the speed of the electronics used to couple modes across large frequency bands.

While the method provided in this section provides a means to construct programmable matrix multiplication operations, it requires that the pulses have a Gaussian envelope with a long duration. Furthermore, our design necessitates the use of multiple cascaded rings to ensure full expressivity of each layer. These pulses undergo FWM in the ring, and propagate through consecutive microring resonators. To overcome these constraints, one could replace the series of microring resonators, with a single ring, that actively captures the neurons, and then stores them for sufficiently long to perform the FWM operations. In such a setup we can also arbitrarily increase the speed of processing by scaling up the strength of the pumps. This "active coupling" approach is presented in the next section and is the one considered for the rest of the manuscript.  

\subsection{Method of Active Coupling}

\begin{figure*}
\centering
\includegraphics[width = \textwidth]{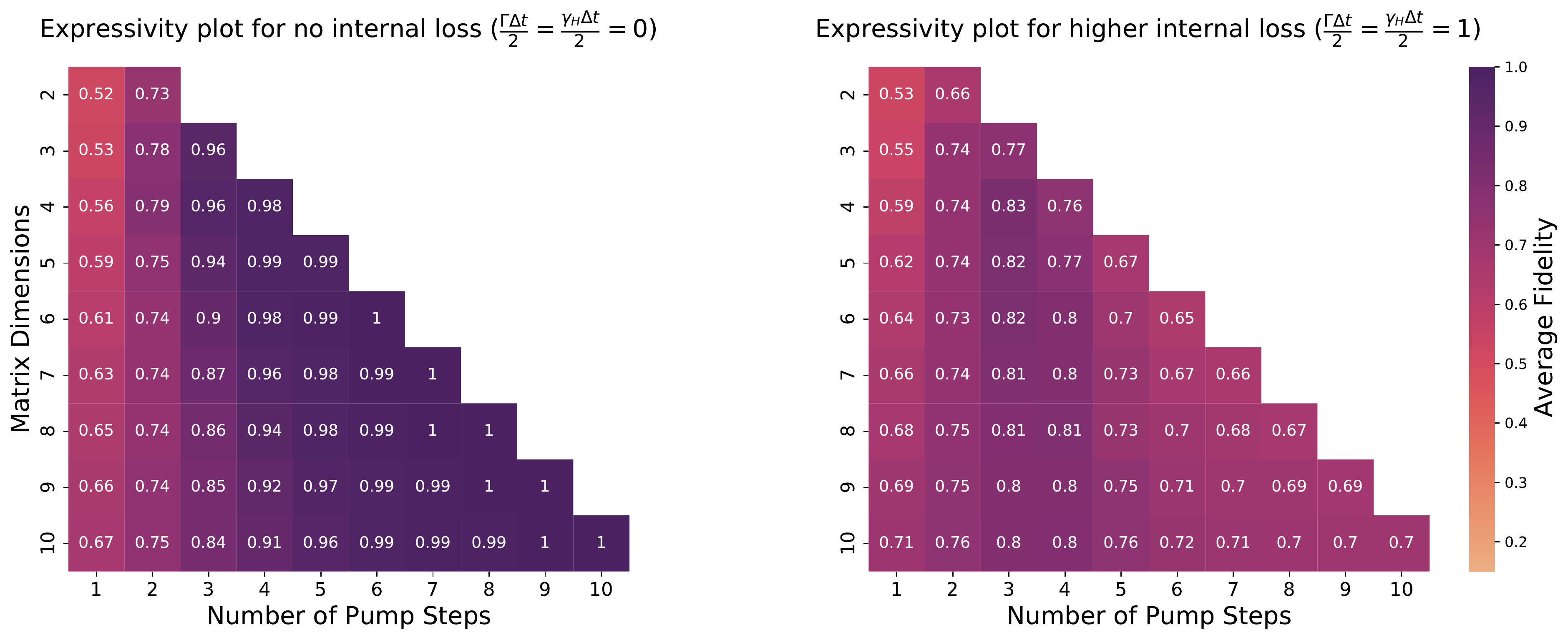}
\caption{ \label{Active_Express}The expressivity of the \emph{active coupling} transformation of the form $\prod e^{\Delta t \textbf{P}}$ in different parameter regimes. Each plot displays the average fidelity as we vary the number of sub-layers, i.e., number of timesteps (the horizontal axis) for a given matrix dimension (the vertical axis).  On the left, the ideal case of no internal loss ($\Gamma = \gamma_\mathrm{H} = 0$) where the expressivity reaches unity at sufficiently many sub-layers/timesteps. On the right, the expressivity at higher loss ($\frac{\Gamma \Delta t}{2} = \frac{\gamma_\mathrm{H} \Delta t}{2} = 1$) never approaches unity. Of note is that the expressivity initially climbs up with the number of layers, until the loss becomes too significant, exponentially growing with the number of layers. Importantly, the expressivity as defined here includes both infidelity due to missing degrees of freedom, and amplitude decay due to leakage from the cavity.}
\end{figure*}

In this section, we propose an architecture with lower circuit size requirements, by using a single microring resonator, as opposed to the linearly-growing sequence of cascaded resonators discussed in the previous section. The neuron activations are still encoded in the complex amplitudes of the neuron frequency states and the linear transformations are encoded in the amplitudes of the pump modes. However, the neuron modes are to be captured and stored in the resonators for the entirety of the FWM process, as opposed to flying by. Such active coupling would require controllable couplings $\gamma(t)$ between the ring and waveguide. We will be constrained by the Q-factor of the ring resonator, giving us an upper bound on how long we can operate on neuron modes before information loss -- we consider methods to overcome that limitation below.

In this design, the pump modes are time-dependent in order to enable full "expressivity" over the neural modes, i.e. the application of any unitary operation. This time dependence is a "continuous" analog to the set of cascaded ring resonators in the previous section. We represent the pump amplitudes as piece-wise constant with step duration of $\Delta t$ to simplify our numerical experiments. The ring-waveguide coupling $\gamma$ is controllable in order to permit the active coupling of the neuron modes, as depicted in Fig.~\ref{Ring_BS}. During capture or release $\gamma$ is increased in order to transfer the neuron mode. During the FWM process $\gamma$ is kept at its minimal value to avoid information loss, thus $\Gamma = \gamma+\gamma_\mathrm{H}=\gamma_\mathrm{H}$. The Hamiltonian of the system during FWM is given by Eq.~\eqref{hamiltonian_FWM}, assuming phase-matched modes. For such a ring, as described in the previous section, the coupled amplitude equations for $N$ neuron modes are given by
\begin{equation}
    \dot{P}_{i} (t) = 0 = -\frac{\Gamma}{2}P_{i}(t) - \sqrt{\gamma}S_{\mathrm{in}, P} (t),
    \label{eq:pump_powers_2}
\end{equation}
\begin{multline}
    \frac{\mathrm{d} A_{i}}{\mathrm{d}t} = \left( -\frac{\Gamma}{2} + i\chi|P_{1}|^{2} \right) A_{i} \\ - \chi \left[ \sum_{j>i}^{N} \left( P_{1}P_{j}^{*} \right) A_{j}   - \sum_{j<i}^{i - 1} \left( P_{1}^{*}P_{j} \right) A_{j} \right].
    \label{eq:cap_neurons}
\end{multline}

In terms of matrix-vector operations, Eq.~\eqref{eq:cap_neurons} can be written as $\dot{\vec{A}} = \textbf{P}\vec{A}$, where $\textbf{P}$ is the Toeplitz matrix from Eq.~\eqref{toeplitz}. The solution to this system of equations (at the end of a period $\Delta t$ during which $\textbf{P}$ is constant) is $\vec{A}(t = \Delta t) = e^{\Delta t \textbf{P}} \vec{A}(t = 0)$. While we have assumed piecewise constant $\textbf{P}$ for simplicity in this example, a freely evolving $\textbf{P}$ is just as easy to work with.

Just as in the previous case, this single-timestep solution provides only $N$ degrees of freedom, as opposed to the $\mathcal{O}(N^{2})$ degrees of freedom in a fully-trainable weight matrix. In this case, however, due to the fact that the pumps are time-dependent, we can perform multiple "sub-layers" by varying the values of the pumps in $\Delta t$ timesteps, without exiting and re-entering ring resonators. Thus, after a time of $N \Delta t$, the net transformation would have $N^{2}$ degrees of freedom, increasing the expressivity. We numerically evaluate the expressivity for different values of $\Gamma$, as described in the previous section, in Fig.~\ref{Active_Express}.

In the ideal case ($\Gamma\Delta t = \gamma_\mathrm{H}\Delta t = 0$), we see that the expressivity grows upon cascading sub-layers just as in the previous case, approaching unity. For a much more pessimistic case where $\frac{1}{2}\Gamma\Delta t = 1$, however, there is a high sensitivity to the loss. We observe an increase in the average fidelity upon cascading a few sub-layers, beyond which the expressivity begins to decrease due to the pulses entering the decay regime. In this case, the final expressivity, even after cascading enough layers to obtain $N^{2}$ degrees of freedom does not reach unity.

\begin{figure}
\centering
\includegraphics[width = \columnwidth]{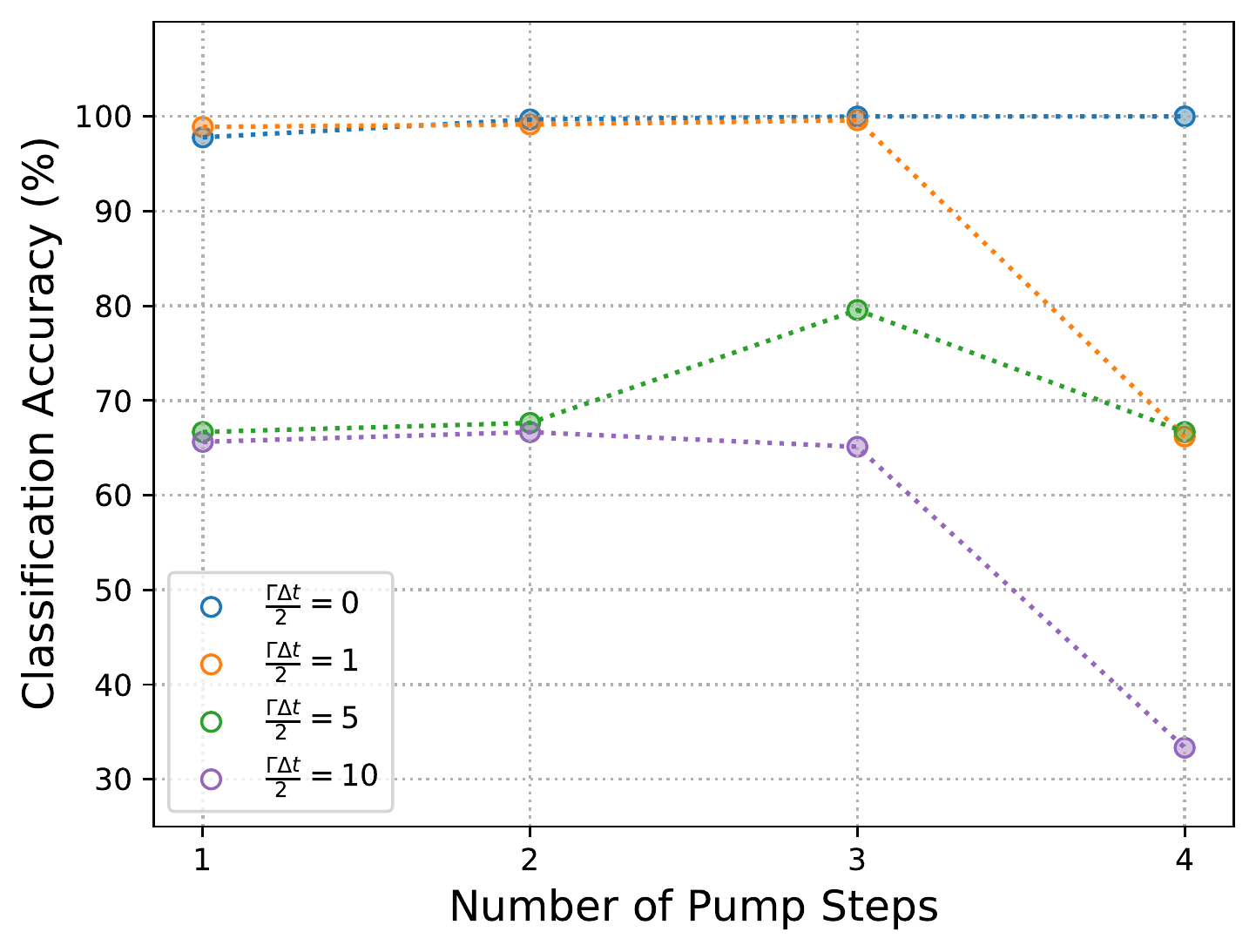}\\
\caption{\label{fig:Iris_active} The classification accuracy on the Iris dataset as we vary the number of cascaded sub-layers (pulse duration) and internal cavity loss rates for the case of \emph{active coupling}. A larger number of pump steps increases then decreases the performance of the network, as expected given the behavior of the expressivity seen in Fig.~\ref{Active_Express}.}
\end{figure}

We illustrate the impact of increasing loss on the classification of the Iris dataset in Fig.~\ref{fig:Iris_active}.
We use an architecture similar to the one described in the previous section, with only a single layer, varying the pulse duration from $\Delta t$ (single step) to $4\Delta t$ (four piece-wise constant steps). For fairly small values of $\Gamma\Delta t$, the network performs well in the task of linear classification, giving us an upward of 99\% accuracy. Increasing $\Gamma\Delta t$, however, increases the classification accuracy only at first, due to the increase in expressivity as we see from Fig.~\ref{Active_Express}. Cascading more than 3 sub-layers increases the losses too much leading to a decrease in overall expressivity, and, correspondingly, the performance of the network. 

\subsection{Computational Speed}

\begin{figure}
    \centering
    \includegraphics[width = \columnwidth]{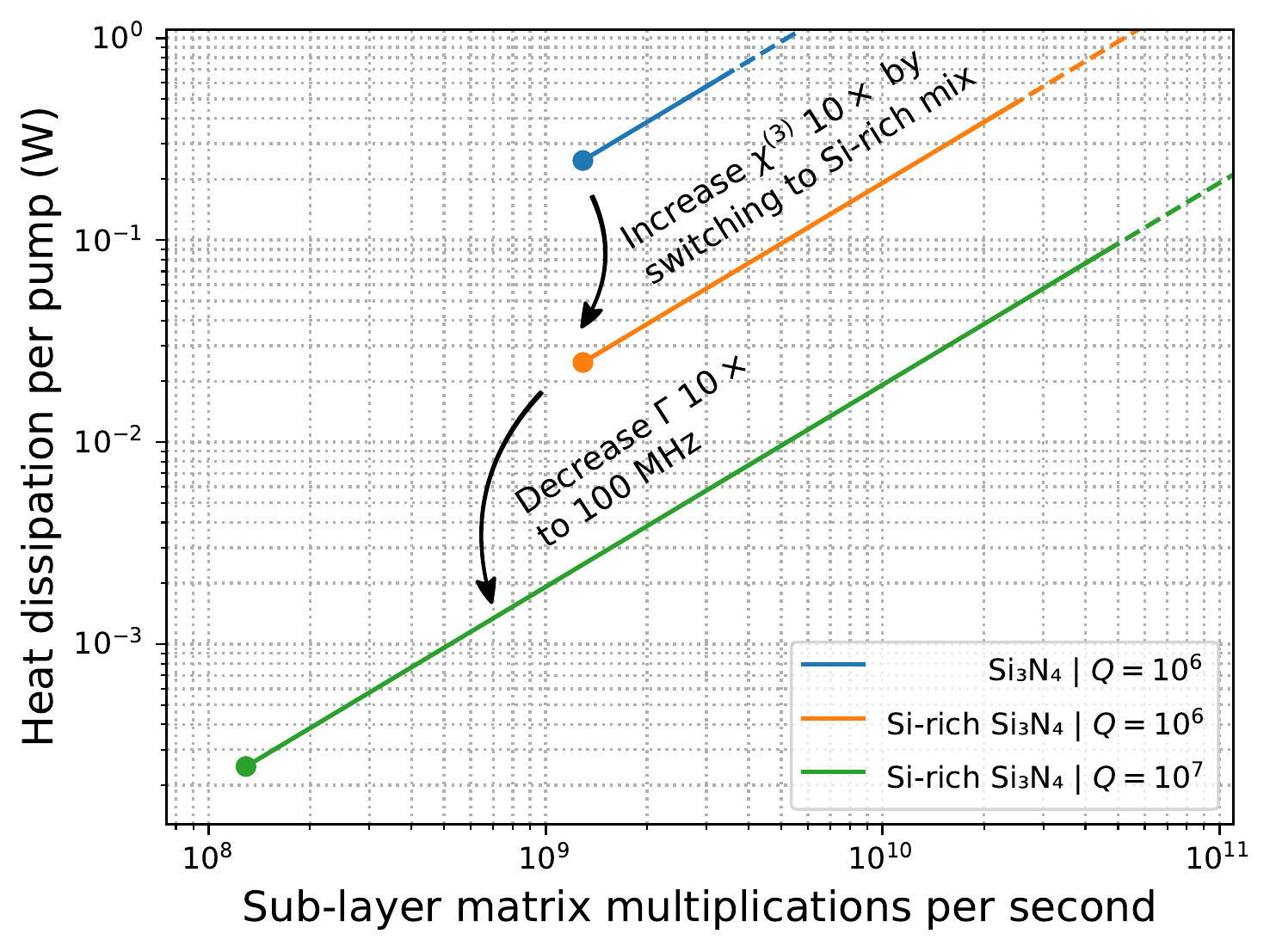}\\
    \caption{Estimated computational performance of our design. The horizontal axis is the rate at which sub-layer multiplications are performed (i.e. the rate of matrix-vector multiplications, where the matrix is a restricted unitary matrix). The vertical axis is an estimate of the heat dissipation expected in a single ring due to leakage from the pump which encodes the matrix parameters. The top blue line represents a typical Silicon Nitride ring~\citep{FWM_ring_1} with $\Gamma=1\mathrm{ns}^{-1}$ and $V_\textrm{FWM}=1300\mathrm{\mu m}$. Two near-term evolutions are presented as well, first (in orange) using Silicon-rich material that significantly increases the $\chi^{(3)}$ susceptibility, and second (in green) developing higher-Q resonators. Lowering the mode volume of the ring would provide similar performance improvements. The curves are cut off to the left due to the constraint seen in Fig.~\ref{Active_Express} and Fig.~\ref{fig:Iris_active} that the computational rate is faster than the decay rate.}
    \label{fig:summary_and_improvements}
\end{figure}

As we have seen in the previous section, the rate at which the wave-mixing interactions happen scales as $\chi P'P''$, where $P'$ and $P''$ denote the pump amplitudes of the main pump and an arbitrary secondary pump. Therefore, the higher the pump power is, the faster the computation can be executed, up to loading and heating constraints. The value for $\chi$ for a given piece of hardware is derived below, giving us realistic engineering constraints on the computational speed. From~\cite{q_nl}, we see that the nonlinear component of the Hamiltonian is given by
\begin{equation}
    \hat{H} = \int \frac{\chi^{(3)}\hat{\textbf{D}}^{4}}{4 \varepsilon_{0}^{3} \eta^{8}} \mathrm{d} \textbf{r},
\end{equation}
where $\chi^{(3)}$ is the FWM nonlinear susceptibility of the material, $\varepsilon_0$ and $\eta$ are the vacuum permitivity and refractive index of the material, and $\hat{\textbf{D}}$ is the electrical displacement field operator. The field operator $\hat{\textbf{D}}$ is the sum of pump or neuron modes $\hat{m}$ that can be written in terms of the eigenmode $\mathbf{d}(\mathbf{r})$ as \cite{q_nl}. 
\begin{equation}
    \hat{\textbf{D}}_{m}(\textbf{r}) = \sqrt{\frac{\hbar \omega_{m}}{2}} \hat{m} \textbf{d}_{m}(\textbf{r}) + \text{H.c.},
\end{equation}
where $\hat{m}$ is the creation operator for the given mode and the normalization condition $\int |\mathbf{d}(\mathbf{r})|^{2} \mathrm{d}\mathbf{r} = \varepsilon_{0}\eta^2$ is fulfilled. Taking into account the energy matching conditions for two neuron modes $\hat{a}_1$ and $\hat{a}_2$ and two pump modes $\hat{p}_1$ and $\hat{p}_2$, and identifying with Eq.~\eqref{hamiltonian_FWM} gives us
\begin{equation}
    \hbar \chi = \frac{3}{2}\frac{\chi^{(3)}}{\varepsilon_{0} \eta^{4}V_{\mathrm{FWM}}} \sqrt{\hbar^{4} \omega_{a_{1}} \omega_{a_{2}} \omega_{p_{1}} \omega_{p_{2}}}
    \label{chi_eff},
\end{equation}
where we define the FWM mode volume $V_{\mathrm{FWM}}$ as  
\begin{equation}
    \frac{1}{V_{\mathrm{FWM}}} = \frac{\int_\text{nl} d_{a_{1}}^{i} d_{a_{2}}^{j*} d_{p_{1}}^{k} d_{p_{2}}^{l*} \mathrm{d}\mathbf{r}}{\sqrt{\int |\mathbf{d}_{a_{1}}|^{2} \mathrm{d}\mathbf{r} \int |\mathbf{d}_{a_{2}}|^{2} \mathrm{d}\mathbf{r} \int |\mathbf{d}_{p_{1}}|^{2} \mathrm{d}\mathbf{r} \int |\mathbf{d}_{p_{2}}|^{2} \mathrm{d}\mathbf{r} }}
    \label{V_FWM}.
\end{equation}
The $\int_\text{nl}$ denotes integration over the volume of the nonlinear material and $i, j, k, l$ denote the spacial components of the fields between which nonlinear interaction is enabled.

If we are to use a Silicon Nitride resonator ($\eta=2.02$ and $\chi^{(3)}=\frac{4}{3}\eta^2n_2\varepsilon_0 c\approx 3.5\times10^{-21} \frac{\mathrm{m}^2}{\mathrm{V}^2}$~\citep{rpphotonics}) with good phase matching such that the FMW mode volume $V_\textrm{FWM}$ is comparable to the geometric volume ($\approx 1300~\mathrm{\mu m^{3}}$ for a $150 \mathrm{\mu m}$ radius, $2.5~\mathrm{\mu m}$ width and $0.73~\mathrm{\mu m}$ height)\cite{FWM_ring_2}, we find $\chi\approx 4.2~\text{s}^{-1}$.

The period of complete exchange of energy between two neuron modes can be calculated via the coupled mode equations derived from Eq.~\eqref{hamiltonian_FWM}, leading to $\Delta t = 2\pi / ( \chi \langle P_{1} \rangle \langle P_{2} \rangle)$, where the maximum amplitudes $\langle P_{*} \rangle$ are measured in square root of average number of photons. We will use these amplitudes as a worst-case estimate of the energy requirements for our design. As we have seen from Fig.~\ref{Active_Express} and Fig.~\ref{fig:Iris_active}, increasing $\Delta t \Gamma$ beyond unity significantly decreases the performance of our hardware due to losses, which leads to the requirement $\langle P_{1} \rangle \langle P_{2} \rangle > \frac{2\pi\Gamma}{\chi} $. For a modern Silicon Nitride resonator we can expect a $Q\approx 10^6$ and $\Gamma=\gamma_\mathrm{H}=\frac{\omega}{Q}\approx 1\mathrm{ns}^{-1}$, therefore $\frac{2\pi\Gamma}{\chi}\approx 10^9$. This implies we need on the order of one billion photons in the main pump mode, leading to thermal heating losses from the main pump on the order of $\Gamma\hbar\omega\langle P\rangle^2\approx 100\mathrm{mW}$.

To summarize, increasing the power of the pumps ($\langle P\rangle^2$) would linearly increase the rate at which computations are performed ($\chi\langle P\rangle^2$) and linearly increase the power dissipated during the computation ($\Gamma\hbar\omega\langle P\rangle^2$). For a typical ring resonator today~\cite{FWM_ring_1}, this implies computational speed of $1\mathrm{GHz}$ (1 billion sub-layer matrix multiplications per second) at dissipation from the main pump of $100\mathrm{mW}$. As seen in Fig.~\ref{fig:summary_and_improvements}, both of these figures of merit can be drastically improved in the very near term by employing already demonstrated techniques (higher $\chi^{(3)}$ in slightly more exotic materials like Silicon-rich Silicon Nitride or AlGaAs and better Q factors). Curiously, there is a lower bound for the computational speed of our device: we need to provide enough pump power such that the computation happens faster than the rate of decay of the neuron modes. On the other hand, microring resonators with arbitrarily large Q factors would pose limitations in pulse capture efficiency, due to their inherent decoupling from the local environment. The controllable coupling coefficient $\gamma(t)$ introduced for active coupling bypasses this issue by allowing us to tune the instantaneous coupling between the resonator and waveguide at the cost of requiring more sophisticated control and fabrication.

Interestingly, the speed of performing a single sub-layer is a constant that does not depend on the number of neurons. Moreover, as seen in the various examples of neural networks in this text, the number of sub-layers itself does not scale any worse than linearly with the number of neurons (and frequently is much better), demonstrating additional architectural advantages in our proposal.

While more detailed discussion follows, here we digress to briefly discuss techniques to circumvent the aforementioned losses. We focus on techniques that can be implemented directly in the ring resonator hardware. A large family of approaches exists for amplifying optical signals, in particular through phase-sensitive amplification in a nonlinear interaction with a pump mode. Given that this is exactly the type of dynamics we are already exploiting, we can add "amplification" to the objective function being optimized during training, ensuring that \emph{typical} patterns of neuron activations are amplified. However, at first sight, the dynamical equations we have derived do not permit any amplifications. The operations we realize are at best unitary (energy preserving). This is due to our assumption of strong non-depleted pumps, which was crucial to obtaining a closed form expression for the linear operations executed by the neural network. Without these closed form expressions, training this neural network becomes more challenging, as it requires backpropagation through the numerical solution of a possibly stiff system of differential equations. However, sacrificing ease of training provides a significant advantage as now typical neural activation patterns can be amplified in order to circumvent losses. In the following section, when we discuss non-linear element-wise activation functions between layers of the neural network, we provide additional techniques to circumvent losses that do not have drawbacks in training.

\section{\label{sec:activation}Non-linear Activation}

\begin{figure}
\centering
\includegraphics[width=\columnwidth]{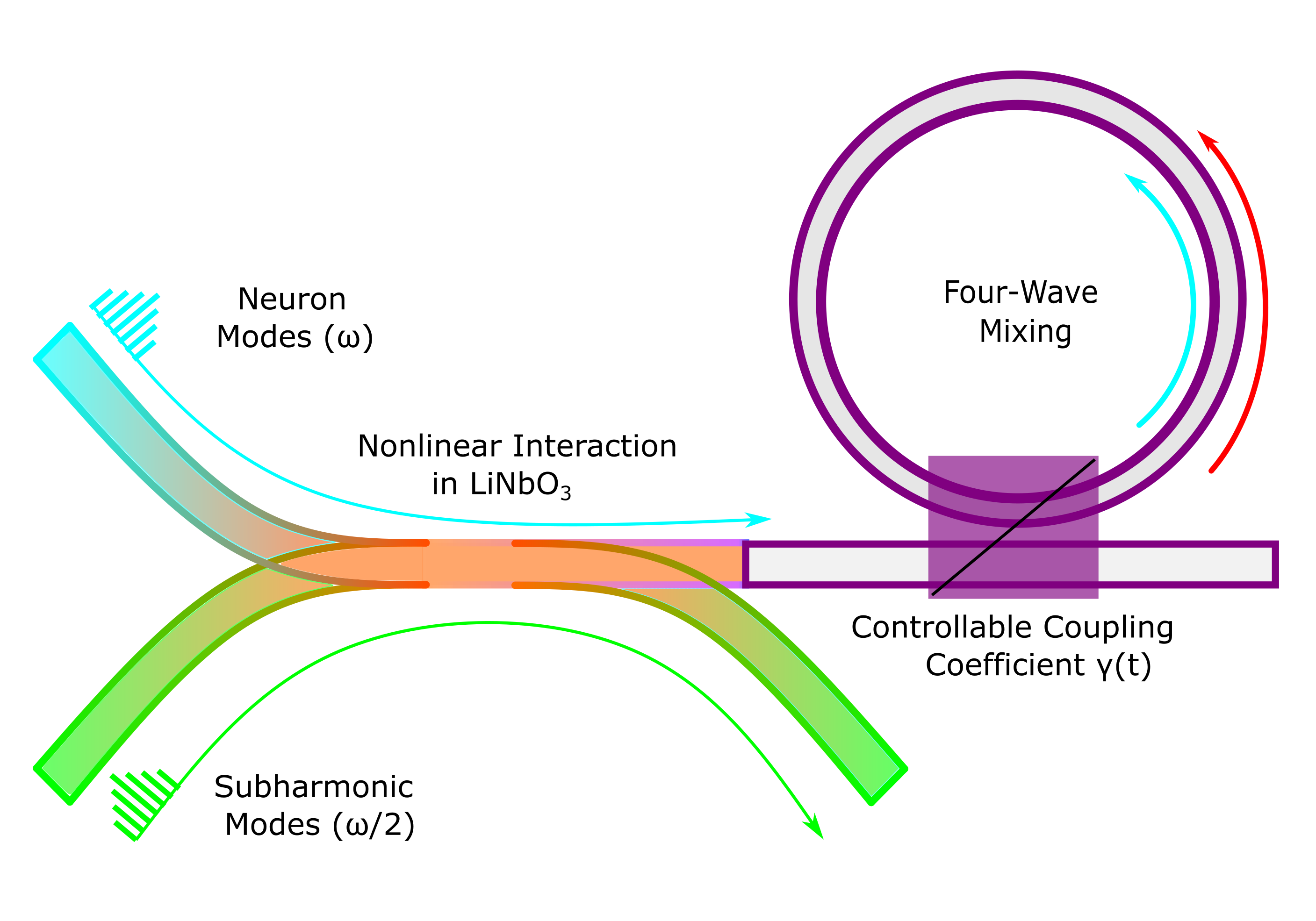}\\
\caption{\label{fig:NL_scheme}  Schematic of the propagating "neural" pulses undergoing the activation function. Input pulses (cyan) are distorted via second order nonlinear interaction in the $\chi^{(2)}$ waveguide before being captured in the ring resonator. The controllable coupling coefficient $\gamma(t)$ allows us to selectively absorb pulses, with efficiency dependent on the how distorted a pulse is.}
\label{Ring_BS}
\end{figure}

\begin{figure*}
    \centering
    \includegraphics[width = \textwidth]{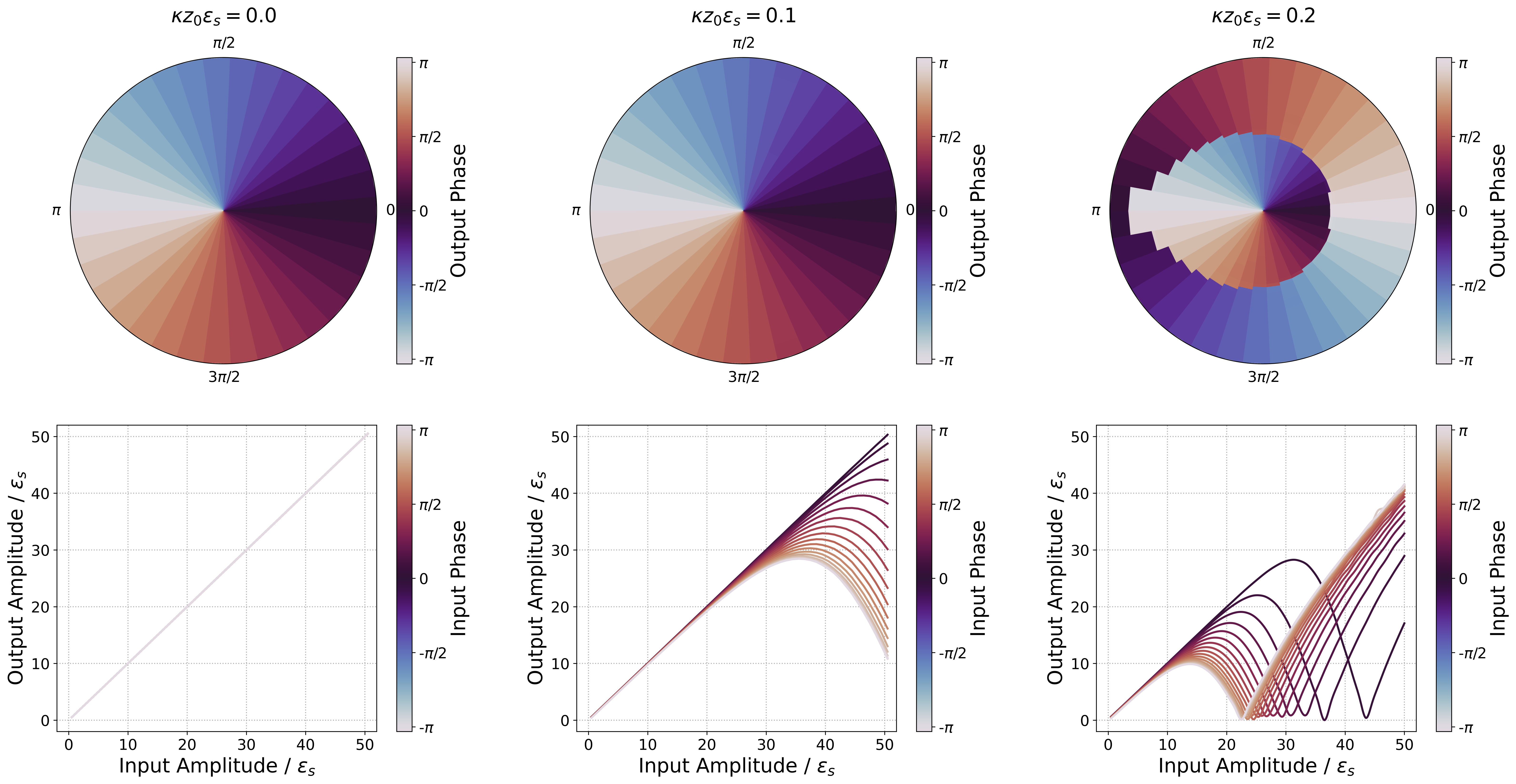}
    \caption{The "neural" activation function realized in our design. The top row of polar plots gives the phase of a neural mode post-activation function (in color) versus the phases of an input mode (polar coordinate) and its amplitude (radial coordinate). We plot the nonlinearity for three different values of the dimensionless parameter $\kappa z_{0} \epsilon_{s} \in \{0.0, 0.1, 0.2\}$. The bottom row presents the output amplitudes (vertical axis) versus the input amplitudes (horizontal axis), scaled to the fixed amplitude of the pump pulses $\epsilon_{s}$. In the absence of nonlinear interaction, i.e., $\kappa z_{0}\epsilon_{s} = 0.0$, we see a linear activation function. The nonlinearity of the activation function becomes more pronounced as the rate of optical nonlinear interactions increases.}
    \label{NL_sub}
\end{figure*}

The nonlinear activation function is indispensable to the operation of the neural network. Previous implementations of the non-linear activation function have relied on the use of thermo-optic effects~\cite{ONN_histories_2, thermo_optic}, hybrid optical-electronic schemes~\cite{nonlinear_acs_1, nonlinear_acs_2}, semiconductor lasers~\cite{nonlinear_acs_5, nonlinear_acs_6} and saturable absorption~\cite{nonlinear_acs_3, nonlinear_acs_4}. The non-linearity we propose relies on nonlinear interactions facilitated by a $\chi^{(2)}$ medium, followed by controllable capture into a ring resonator. However, the linear transformations introduced in the previous sections places constraints on the degree of non-linearity that can be imposed as we show in this section.

The non-linearity we propose is based upon a second-order nonlinear interaction (e.g., in a Lithium Niobate waveguide, characterized by its $\chi^{(2)}$ susceptibility coefficient). We release the neuron mode from the resonator in which the matrix multiplication was performed into the waveguide. We aim to distort the temporal envelope of the neuron mode via the nonlinear interaction with an externally pumped pulse (that we term as the subharmonic mode). This subharmonic mode has a frequency of half of the neuron mode. Following the distortion, we selectively capture~\cite{catch_release_1, catch_release_4} the neuron mode into the microring resonator that forms the subsequent layer of the neural network. Thus, the distorted pulses are selectively absorbed into the ring, with absorption efficiency dependent on the amount of distortion. The non-linear distortion is stronger for higher-amplitude pulses, giving rise to a total effective non-linearity. Fig.~\ref{Ring_BS} provides a sketch of the setup. To avoid interactions between different neural modes, i.e. keep the activation function element-wise, a waveguide segment with dispersion can be used to offset the modes in time. 

First, we explore the envelope distortion dynamics for a neural pulse interacting with a subharmonic pump pulse in a waveguide. We parametrize both envelopes as $E_\mathrm{n}(z,t)$ and $E_\mathrm{sub}(z,t)$ where $z$ is the spacial coordinate along the length of the waveguide. As elaborated in the supplementary materials, these envelopes obey~\cite{PDE} 
\begin{equation}
    \frac{\partial E_{\mathrm{n}}}{\partial z} + \frac{\eta}{c} \frac{\partial E_{\mathrm{n}}}{\partial t} = -\kappa E_{\mathrm{sub}}^{2} - \alpha E_{\mathrm{n}}
    \label{PDE_1}
\end{equation}
\begin{equation}
    \frac{\partial E_{\mathrm{sub}}}{\partial z} + \frac{\eta}{c} \frac{\partial E_{\mathrm{sub}}}{\partial t} = \kappa E_{\mathrm{n}} E_{\mathrm{sub}}^{*} - \alpha E_{\mathrm{sub}}
    \label{PDE_2}
\end{equation}
\begin{equation}
    \kappa = \frac{\omega}{c} \chi^{(2)} s,
\end{equation}
where $s$ is a unitless measure of the mode overlap between the neural and subharmonic modes, $\omega$ is the frequency of the neural mode, and $\alpha$ is the waveguide loss. For specificity, we consider Gaussian wavepackets for the input neural modes (released from the ring that has been performing the matrix multiplication of the previous layer) of the form $E_{\mathrm{n}} = \epsilon_{\mathrm{n}} e^{-\frac{z^2}{4w^2}}e^{-i\varphi_{0}}$, where $w$ is the spacial length of the packet, $\varphi_0$ is the phase of the neuron activation, and $\epsilon_\mathrm{n}$ gives the field amplitude scale. Similarly, for the subharmonic pump we set $E_{\mathrm{sub}} = \epsilon_{\mathrm{s}} e^{-\frac{z^2}{4w^2}}$, however an equally valid option would be a continuous wave $E_{\mathrm{sub}} = \epsilon_{\mathrm{s}}$. We solve for the evolution of $E_\mathrm{n}(z,t)$ numerically. The dimensionless parameters that emerge as chiefly governing these dynamics are the effective strength of the nonlinear interaction $\kappa\epsilon_s z_0$ and the strength of the neural mode relative to the fixed subharmonic mode, $\epsilon_\mathrm{n}/\epsilon_\mathrm{s}$.
The length of the $\chi^{(2)}$ waveguide is denoted $z_0$. The simulations and the necessary algebraic manipulations are detailed in the interactive supplementary materials.

The distorted neuron envelopes are then actively captured into the next ring via a controllable ring-waveguide coupling $\gamma(t)$. The dynamics of the capture without interactions from the pump modes is governed by~\cite{CMT, catch_release_1, catch_release_5, catch_release_6} 
\begin{equation}
    \frac{\mathrm{d}A}{\mathrm{d}t} = -\frac{(\gamma(t) + \gamma_\mathrm{H})}{2}A + \sqrt{\gamma(t)}S_{\mathrm{in}}
    \label{capture_1}
\end{equation}
\begin{equation}
    S_{\mathrm{out}} = S_{\mathrm{in}} + \sqrt{\gamma(t)}A,
    \label{capture_2}
\end{equation}
where $S_\mathrm{in}(t)=E_\mathrm{n}(0,t)$ is the incoming neural mode's envelope, $S_\mathrm{out}$ is the outgoing (not captured) signal, and $A$ is the neuron mode amplitude captured in the resonator. By fixing $S_\mathrm{out}=0$ we can solve for the $\gamma(t)$ that would completely capture a given envelope $S_\mathrm{in}$. The analytical solution for a Gaussian wave packet is given in the supplementary materials. However, high neural activations would lead to strong envelope distortions, which in turn cause the mode to not be fully captured, thus providing for the equivalent of a non-linear element-wise activation function in our NN architecture. Importantly, this implementation naturally supports negative activations, unlike the vast majority of optical approaches. Arbitrary complex values are supported as can be seen in Fig.~\ref{NL_sub}.

\begin{figure*}
\includegraphics[width = \textwidth]{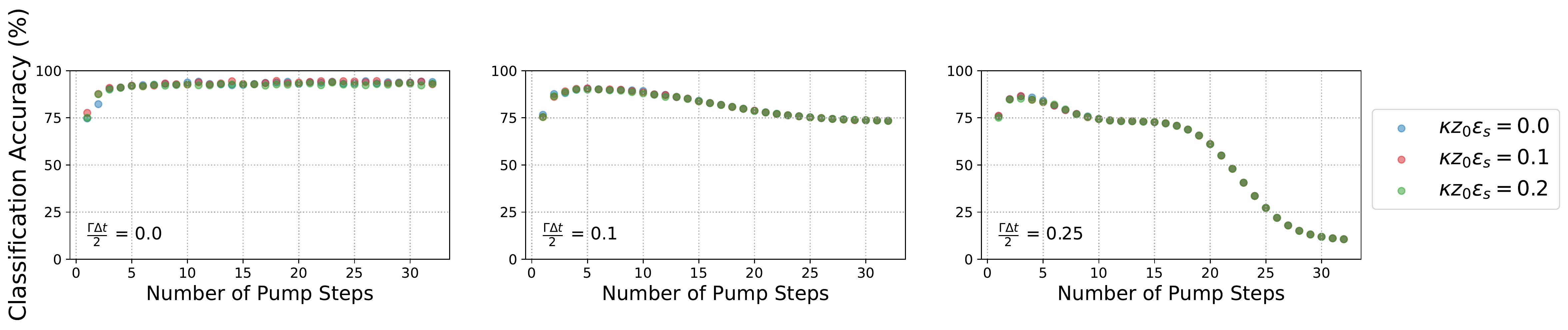}
\includegraphics[width = \textwidth]{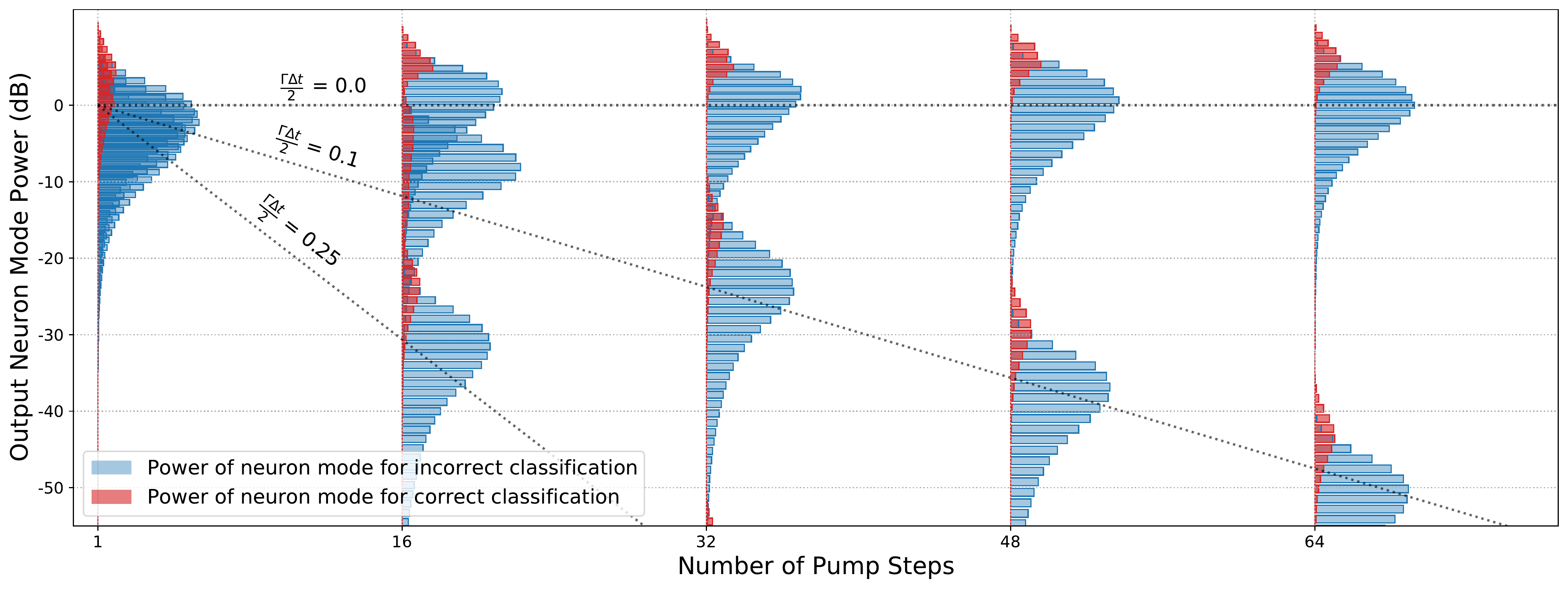}
\caption{ \label{MNIST_decay} The classification performance of an all-optical neural net against the MNIST dataset depending on optical losses, effective waveguide nonlinearity, and network size. We use varying number of 64-neuron sublayers (as depicted on the horizontal axis) followed by ten 10-neuron layers. In the top row we have classification accuracy (vertical axis) versus number of sub-layers, i.e., distinct piece-wise constant steps in the control pumps (horizontal axis). The three top facets depict different decay rates $\Gamma$, e.g., $\frac{\Gamma\Delta t}{2}=0.25$ corresponds to $\Gamma=0.5\mathrm{ns}^{-1}$ for a step duration of $\Delta t = 1\mathrm{ns}$. The strength of the nonlinear interaction in the waveguides between ring-resonators is depicted in the color of the marker. While, initially, increasing the number of sub-layers improves the performance thanks to the higher expressivity of the encoded operation, further increase is detrimental as it causes non-unitary behavior and decrease in expressivity. In the third facet one can additionally observe the precipitous drop in performance when, due to the increasing losses, the shot noise starts dominating the measurement result.
The bottom plot shows histograms of the power carried by the output neural modes. For various number of sub-layers (horizontal axis) and various loss rates (annotated with dashed lines) we plot the distribution of energy per neuron mode (histograms with respect to the vertical axis). The blue histograms correspond to the "incorrect class" neurons, while the red are the "correct class" neurons (which are fewer). The energy carried by the "correct class" neurons is consistently higher, indicating effective classicafication. Moreover, at high expressivity the red histogram has noticeably smaller spread. Lastly, we can see the rapid loss of power as the number of time steps grows in the presence of loss, explaining the precipitous drop in performance seen in the top-right plot.
}
\end{figure*}

From the numerical experiments we see that $\kappa\epsilon_s z_0\approx0.2$ provides for a saturating activation function. For a waveguide of length $z_0=1\mathrm{cm}$, with good mode overlap $s\approx1$, in Lithium Niobate with $\chi^{(2)}=31\mathrm{\frac{pm}{V}}$ we obtain $\epsilon_s=160\mathrm{\frac{kV}{m}}$. Such a field strength amplitude corresponds to a peak power of approximately $\varepsilon_0\sqrt\eta c\epsilon_s^2 a=20\mathrm{\mu W}$ for a waveguide with cross-section of $a=0.2\mathrm{\mu m}^2$. Such pump powers are easy to achieve and should pose no problem for the realization of our device. Depending on the platform, especially if one wants to avoid heterogenous integration, other materials with a high $\chi^{(2)}$, Galium Arsenide~\cite{GaAs}, Aluminium Gallium Arsenide~\cite{AlGaAs}, and Silicon Carbide~\cite{SiC} can be used. Such materials with both high $\chi^{(2)}$ and $\chi^{(3)}$ coefficient would allow the entire device to be integrated into a single material platform~\cite{Single_platform_1, Single_platform_2}.

The non-linear activation function can also be used to circumvent losses experienced in the ring resonators. The non-linear interaction can be engineered to provide an activation function with a slope greater than one, instead of a sigmoid-like function. Examples are discussed in the supplementary materials.

\section{\label{sec:casestudy}Case Study: Image Classification}

We benchmark the performance of the proposed hardware designs in a simulated neural network for image classification. We classify the MNIST dataset of handwritten digits. Our training set consists of 50000 $28 \times 28$-pixel images. 

The optical neural network is trained on the low-frequency Fourier features of the MNIST images. To do this, we pre-processes the images by truncating the central $N \times N$ window from the two-dimensional Fourier transform of the images. These bitmaps are then reshaped into vectors of size $N^{2}$. These vectors are encoded into the initial complex amplitudes of the modes of the simulated microring resonator, i.e., the input layer of the neural network. We choose $N = 8$ as the low-frequency components contain most of the pertinent information about the image.
Our training uses mini-batch gradient descent for 200 epochs with the Adam optimizer~\cite{Adam}. Each batch contains 2000 images of the training data. The learning rate decays exponentially, from 0.01 at the start to 0.0002 at the end.

The linear transformations are implemented through the method of "active coupling".
In our simulations, we test the performance of the network in different loss regimes while varying the number of sub-layers, i.e. the piece-wise constant steps of the pumps.
In previous sections we have observed that expressivity initially grows with the number of sub-layers, until losses due to the prolonged operations become detrimental. This observation is confirmed in Fig.~\ref{MNIST_decay} where we show the classification accuracy of our model versus the various hardware parameters: some minimum number of sub-layers is required to reach sufficiently good accuracy, after which accuracy degrades due to losses. To explicitly illustrate this loss regime we also plot histograms of the energy carried by each neuron mode at the output of the neural network. Unsurprisingly, the energy of the neurons decays exponentially as we increase the number of sub-layers. Even before measurements become shot-noise limited, the performance of the network drops. For a state-of-the art cavity ($\Gamma=0.2 \mathrm{ns}^{-1}$) and control pulse resolution of $\Delta t=1\mathrm{ns}$ we obtain excellent classification performance and less than 5 dB of loss. However, for larger networks, the pumping schemes discussed in previous sections would be crucial for the reliable performance of the system.

\section{Discussion and Conclusion}

This work presents a novel architecture for an all-optical artificial neural network processor that relies solely on non-linear optical processes. The proposed scheme encodes information in the complex amplitudes of frequency states and are modulated via four-wave mixing in a $\chi^{(3)}$ medium, that enables the process of matrix multiplication. The non-linear activation function relies on the $\chi^{(2)}$ interactions with subharmonic modes resulting in distortion of the neuron mode pulses and consecutive projection on a Gaussian pulse shape. Furthermore, the scheme we propose can be realized experimentally on-chip with ease, via only microring resonators. 

The proposed neural network processor has multiple advantages over previous implementations of optical and electronics neural networks. As opposed to digital matrix-vector multiplication that typically scales as $O(N^{2})$,  the proposed model has a time complexity of only $O(N)$, due to the fact that for any given instance of FWM, all of the neuron modes are modulated at the same time. The number of on-chip components is also very low, as all neural modes occupy the same resonator. Another feature of the architecture we propose is that the speed of the operations is directly proportional to the power of the pumps, letting us freely increase the computational speed. At extreme speeds this leads to increased heating due to leaking from the pumps, however increases resonator quality can offset this problem. As seen in the main text, with near-term photonic technology one should be able to demonstrate billions of matrix multiplications per second at tens of mW dissipation rates. 

Fruitful directions for future work include extensions of the features of the architecture we propose to a variety of problems, including recursive neural networks for Ising Machines. Moreover, a constructive direction to investigate is the development of in-situ, on-chip training~\cite{backprop} or self-learning machines, as opposed to using numerically optimized pumps. The computations being performed belong to the unitary group, which also opens up many future avenues for ultra-fast reversible computing.

\begin{acknowledgments}
We thank Ryan Hamerly for discussions on the feasibility of the nonlinear activation function.
The authors acknowledge financial support from the MITRE Quantum Moonshot Program. The simulations presented in this paper were performed on the MIT-PSFC partition of the Engaging cluster at the MGHPCC facility (\url{www.mghpcc.org}) which was funded by DoE grant number DE-FG02-91-ER54109. We would like to thank the Tensorflow and Julia open source communities for the research software provided that enabled us to perform simulations in optimization, machine learning, and solving PDEs. M. H. acknowledges funding from Villum Fonden (QNET-NODES grant no. 37417). 
\end{acknowledgments}

\bibliography{main.bib}

\begin{thebibliography}{59}%
\makeatletter
\providecommand \@ifxundefined [1]{%
 \@ifx{#1\undefined}
}%
\providecommand \@ifnum [1]{%
 \ifnum #1\expandafter \@firstoftwo
 \else \expandafter \@secondoftwo
 \fi
}%
\providecommand \@ifx [1]{%
 \ifx #1\expandafter \@firstoftwo
 \else \expandafter \@secondoftwo
 \fi
}%
\providecommand \natexlab [1]{#1}%
\providecommand \enquote  [1]{``#1''}%
\providecommand \bibnamefont  [1]{#1}%
\providecommand \bibfnamefont [1]{#1}%
\providecommand \citenamefont [1]{#1}%
\providecommand \href@noop [0]{\@secondoftwo}%
\providecommand \href [0]{\begingroup \@sanitize@url \@href}%
\providecommand \@href[1]{\@@startlink{#1}\@@href}%
\providecommand \@@href[1]{\endgroup#1\@@endlink}%
\providecommand \@sanitize@url [0]{\catcode `\\12\catcode `\$12\catcode
  `\&12\catcode `\#12\catcode `\^12\catcode `\_12\catcode `\%12\relax}%
\providecommand \@@startlink[1]{}%
\providecommand \@@endlink[0]{}%
\providecommand \url  [0]{\begingroup\@sanitize@url \@url }%
\providecommand \@url [1]{\endgroup\@href {#1}{\urlprefix }}%
\providecommand \urlprefix  [0]{URL }%
\providecommand \Eprint [0]{\href }%
\providecommand \doibase [0]{https://doi.org/}%
\providecommand \selectlanguage [0]{\@gobble}%
\providecommand \bibinfo  [0]{\@secondoftwo}%
\providecommand \bibfield  [0]{\@secondoftwo}%
\providecommand \translation [1]{[#1]}%
\providecommand \BibitemOpen [0]{}%
\providecommand \bibitemStop [0]{}%
\providecommand \bibitemNoStop [0]{.\EOS\space}%
\providecommand \EOS [0]{\spacefactor3000\relax}%
\providecommand \BibitemShut  [1]{\csname bibitem#1\endcsname}%
\let\auto@bib@innerbib\@empty
\bibitem [{\citenamefont {{Young}}\ \emph {et~al.}(2018)\citenamefont
  {{Young}}, \citenamefont {{Hazarika}}, \citenamefont {{Poria}},\ and\
  \citenamefont {{Cambria}}}]{NLP}%
  \BibitemOpen
  \bibfield  {author} {\bibinfo {author} {\bibfnamefont {T.}~\bibnamefont
  {{Young}}}, \bibinfo {author} {\bibfnamefont {D.}~\bibnamefont {{Hazarika}}},
  \bibinfo {author} {\bibfnamefont {S.}~\bibnamefont {{Poria}}},\ and\ \bibinfo
  {author} {\bibfnamefont {E.}~\bibnamefont {{Cambria}}},\ }\href
  {https://doi.org/10.1109/MCI.2018.2840738} {\bibfield  {journal} {\bibinfo
  {journal} {IEEE Computational Intelligence Magazine}\ }\textbf {\bibinfo
  {volume} {13}},\ \bibinfo {pages} {55} (\bibinfo {year} {2018})}\BibitemShut
  {NoStop}%
\bibitem [{\citenamefont {{Senior}}\ \emph {et~al.}(2020)\citenamefont
  {{Senior}}, \citenamefont {{Evans}},\ and\ \citenamefont {et~al.}}]{protein}%
  \BibitemOpen
  \bibfield  {author} {\bibinfo {author} {\bibfnamefont {A.~W.}\ \bibnamefont
  {{Senior}}}, \bibinfo {author} {\bibfnamefont {R.}~\bibnamefont {{Evans}}},\
  and\ \bibinfo {author} {\bibfnamefont {J.~J.}\ \bibnamefont {et~al.}},\
  }\href {https://doi.org/10.1038/s41586-019-1923-7} {\bibfield  {journal}
  {\bibinfo  {journal} {Nature}\ }\textbf {\bibinfo {volume} {577}},\ \bibinfo
  {pages} {706} (\bibinfo {year} {2020})}\BibitemShut {NoStop}%
\bibitem [{\citenamefont {{Chrittwieser}}\ \emph {et~al.}(2020)\citenamefont
  {{Chrittwieser}}, \citenamefont {{Antonoglou}},\ and\ \citenamefont
  {et~al.}}]{muzero}%
  \BibitemOpen
  \bibfield  {author} {\bibinfo {author} {\bibfnamefont {J.}~\bibnamefont
  {{Chrittwieser}}}, \bibinfo {author} {\bibfnamefont {I.}~\bibnamefont
  {{Antonoglou}}},\ and\ \bibinfo {author} {\bibfnamefont {T.~S.~H.}\
  \bibnamefont {et~al.}},\ }\href {https://doi.org/10.1038/s41586-020-03051-4}
  {\bibfield  {journal} {\bibinfo  {journal} {Nature}\ }\textbf {\bibinfo
  {volume} {588}},\ \bibinfo {pages} {604} (\bibinfo {year}
  {2020})}\BibitemShut {NoStop}%
\bibitem [{\citenamefont {{Steinkraus}}\ \emph {et~al.}(2005)\citenamefont
  {{Steinkraus}}, \citenamefont {{Buck}},\ and\ \citenamefont
  {{Simard}}}]{GPU}%
  \BibitemOpen
  \bibfield  {author} {\bibinfo {author} {\bibfnamefont {D.}~\bibnamefont
  {{Steinkraus}}}, \bibinfo {author} {\bibfnamefont {I.}~\bibnamefont
  {{Buck}}},\ and\ \bibinfo {author} {\bibfnamefont {P.~Y.}\ \bibnamefont
  {{Simard}}},\ }in\ \href {https://doi.org/10.1109/ICDAR.2005.251} {\emph
  {\bibinfo {booktitle} {Eighth International Conference on Document Analysis
  and Recognition (ICDAR'05)}}}\ (\bibinfo {year} {2005})\ pp.\ \bibinfo
  {pages} {1115--1120 Vol. 2}\BibitemShut {NoStop}%
\bibitem [{\citenamefont {{Graves}}\ \emph {et~al.}(2016)\citenamefont
  {{Graves}}, \citenamefont {{Wayne}},\ and\ \citenamefont {et~al.}}]{TPU}%
  \BibitemOpen
  \bibfield  {author} {\bibinfo {author} {\bibfnamefont {A.}~\bibnamefont
  {{Graves}}}, \bibinfo {author} {\bibfnamefont {G.}~\bibnamefont {{Wayne}}},\
  and\ \bibinfo {author} {\bibfnamefont {G.~M.~R.}\ \bibnamefont {et~al.}},\
  }\href {https://doi.org/10.1038/nature20101} {\bibfield  {journal} {\bibinfo
  {journal} {Nature}\ }\textbf {\bibinfo {volume} {538}},\ \bibinfo {pages}
  {471} (\bibinfo {year} {2016})}\BibitemShut {NoStop}%
\bibitem [{\citenamefont {Esser}\ \emph {et~al.}(t 11)\citenamefont {Esser},
  \citenamefont {Merolla}, \citenamefont {Arthur}, \citenamefont {Cassidy},
  \citenamefont {Appuswamy}, \citenamefont {Andreopoulos}, \citenamefont
  {Berg}, \citenamefont {McKinstry}, \citenamefont {Melano}, \citenamefont
  {Barch}, \citenamefont {di~Nolfo}, \citenamefont {Datta}, \citenamefont
  {Amir}, \citenamefont {Taba},\ and\ \citenamefont {Flickner}}]{IBM}%
  \BibitemOpen
  \bibfield  {author} {\bibinfo {author} {\bibfnamefont {S.~K.}\ \bibnamefont
  {Esser}}, \bibinfo {author} {\bibfnamefont {P.~A.}\ \bibnamefont {Merolla}},
  \bibinfo {author} {\bibfnamefont {J.~V.}\ \bibnamefont {Arthur}}, \bibinfo
  {author} {\bibfnamefont {A.~S.}\ \bibnamefont {Cassidy}}, \bibinfo {author}
  {\bibfnamefont {R.}~\bibnamefont {Appuswamy}}, \bibinfo {author}
  {\bibfnamefont {A.}~\bibnamefont {Andreopoulos}}, \bibinfo {author}
  {\bibfnamefont {D.~J.}\ \bibnamefont {Berg}}, \bibinfo {author}
  {\bibfnamefont {J.~L.}\ \bibnamefont {McKinstry}}, \bibinfo {author}
  {\bibfnamefont {T.}~\bibnamefont {Melano}}, \bibinfo {author} {\bibfnamefont
  {D.~R.}\ \bibnamefont {Barch}}, \bibinfo {author} {\bibfnamefont
  {C.}~\bibnamefont {di~Nolfo}}, \bibinfo {author} {\bibfnamefont
  {P.}~\bibnamefont {Datta}}, \bibinfo {author} {\bibfnamefont
  {A.}~\bibnamefont {Amir}}, \bibinfo {author} {\bibfnamefont {B.}~\bibnamefont
  {Taba}},\ and\ \bibinfo {author} {\bibfnamefont {D.~S.}\ \bibnamefont
  {Flickner}, \bibfnamefont {M.~D.~Modha}},\ }in\ \href
  {https://doi.org/10.1073/pnas.1604850113} {\emph {\bibinfo {booktitle}
  {Proceedings of the National Academy of Sciences}}},\ Vol.\ \bibinfo {volume}
  {113}\ (\bibinfo {year} {2016 Oct 11})\ pp.\ \bibinfo {pages}
  {11441--1146}\BibitemShut {NoStop}%
\bibitem [{Int(2016)}]{Intel}%
  \BibitemOpen
  \href {https://www.intel.ai/nervana-nnp/} {\bibinfo {title} {Intel nervana
  neural network processor}} (\bibinfo {year} {2016}),\ \bibinfo {note}
  {(accessed: 12.01.2020)}\BibitemShut {NoStop}%
\bibitem [{\citenamefont {Almeida}\ \emph {et~al.}(2004)\citenamefont
  {Almeida}, \citenamefont {Barrios},\ and\ \citenamefont {Panepucci}}]{SiP}%
  \BibitemOpen
  \bibfield  {author} {\bibinfo {author} {\bibfnamefont {V.}~\bibnamefont
  {Almeida}}, \bibinfo {author} {\bibfnamefont {C.}~\bibnamefont {Barrios}},\
  and\ \bibinfo {author} {\bibfnamefont {R.~e.~a.}\ \bibnamefont {Panepucci}},\
  }\href {https://doi.org/10.1038/nature02921} {\bibfield  {journal} {\bibinfo
  {journal} {Nature}\ }\textbf {\bibinfo {volume} {431}},\ \bibinfo {pages}
  {1081} (\bibinfo {year} {2004})}\BibitemShut {NoStop}%
\bibitem [{\citenamefont {Leuthold}\ \emph {et~al.}(2010)\citenamefont
  {Leuthold}, \citenamefont {Koos},\ and\ \citenamefont {Freude}}]{SiP_2}%
  \BibitemOpen
  \bibfield  {author} {\bibinfo {author} {\bibfnamefont {J.}~\bibnamefont
  {Leuthold}}, \bibinfo {author} {\bibfnamefont {C.}~\bibnamefont {Koos}},\
  and\ \bibinfo {author} {\bibfnamefont {W.}~\bibnamefont {Freude}},\ }\href
  {https://doi.org/10.1038/nphoton.2010.185} {\bibfield  {journal} {\bibinfo
  {journal} {Nature photonics}\ }\textbf {\bibinfo {volume} {4}},\ \bibinfo
  {pages} {535} (\bibinfo {year} {2010})}\BibitemShut {NoStop}%
\bibitem [{\citenamefont {{Tait}}\ \emph {et~al.}(2014)\citenamefont {{Tait}},
  \citenamefont {{Nahmias}}, \citenamefont {{Shastri}},\ and\ \citenamefont
  {{Prucnal}}}]{spike}%
  \BibitemOpen
  \bibfield  {author} {\bibinfo {author} {\bibfnamefont {A.~N.}\ \bibnamefont
  {{Tait}}}, \bibinfo {author} {\bibfnamefont {M.~A.}\ \bibnamefont
  {{Nahmias}}}, \bibinfo {author} {\bibfnamefont {B.~J.}\ \bibnamefont
  {{Shastri}}},\ and\ \bibinfo {author} {\bibfnamefont {P.~R.}\ \bibnamefont
  {{Prucnal}}},\ }\href {https://doi.org/10.1109/JLT.2014.2345652} {\bibfield
  {journal} {\bibinfo  {journal} {Journal of Lightwave Technology}\ }\textbf
  {\bibinfo {volume} {32}},\ \bibinfo {pages} {4029} (\bibinfo {year}
  {2014})}\BibitemShut {NoStop}%
\bibitem [{\citenamefont {Vandoorne}\ \emph {et~al.}(2008)\citenamefont
  {Vandoorne}, \citenamefont {Dierckx}, \citenamefont {Schrauwen},
  \citenamefont {Verstraeten}, \citenamefont {Baets}, \citenamefont
  {Bienstman},\ and\ \citenamefont {Van~Campenhout}}]{reservoir}%
  \BibitemOpen
  \bibfield  {author} {\bibinfo {author} {\bibfnamefont {K.}~\bibnamefont
  {Vandoorne}}, \bibinfo {author} {\bibfnamefont {W.}~\bibnamefont {Dierckx}},
  \bibinfo {author} {\bibfnamefont {B.}~\bibnamefont {Schrauwen}}, \bibinfo
  {author} {\bibfnamefont {D.}~\bibnamefont {Verstraeten}}, \bibinfo {author}
  {\bibfnamefont {R.}~\bibnamefont {Baets}}, \bibinfo {author} {\bibfnamefont
  {P.}~\bibnamefont {Bienstman}},\ and\ \bibinfo {author} {\bibfnamefont
  {J.}~\bibnamefont {Van~Campenhout}},\ }\href
  {https://doi.org/10.1364/OE.16.011182} {\bibfield  {journal} {\bibinfo
  {journal} {Opt. Express}\ }\textbf {\bibinfo {volume} {16}},\ \bibinfo
  {pages} {11182} (\bibinfo {year} {2008})}\BibitemShut {NoStop}%
\bibitem [{\citenamefont {{Sze}}\ \emph {et~al.}(2017)\citenamefont {{Sze}},
  \citenamefont {{Chen}}, \citenamefont {{Yang}},\ and\ \citenamefont
  {{Emer}}}]{expend_1}%
  \BibitemOpen
  \bibfield  {author} {\bibinfo {author} {\bibfnamefont {V.}~\bibnamefont
  {{Sze}}}, \bibinfo {author} {\bibfnamefont {Y.}~\bibnamefont {{Chen}}},
  \bibinfo {author} {\bibfnamefont {T.}~\bibnamefont {{Yang}}},\ and\ \bibinfo
  {author} {\bibfnamefont {J.~S.}\ \bibnamefont {{Emer}}},\ }\href
  {https://doi.org/10.1109/JPROC.2017.2761740} {\bibfield  {journal} {\bibinfo
  {journal} {Proceedings of the IEEE}\ }\textbf {\bibinfo {volume} {105}},\
  \bibinfo {pages} {2295} (\bibinfo {year} {2017})}\BibitemShut {NoStop}%
\bibitem [{\citenamefont {Chen}\ \emph {et~al.}(2014)\citenamefont {Chen},
  \citenamefont {Du}, \citenamefont {Sun}, \citenamefont {Wang}, \citenamefont
  {Wu}, \citenamefont {Chen},\ and\ \citenamefont {Temam}}]{expend_2}%
  \BibitemOpen
  \bibfield  {author} {\bibinfo {author} {\bibfnamefont {T.}~\bibnamefont
  {Chen}}, \bibinfo {author} {\bibfnamefont {Z.}~\bibnamefont {Du}}, \bibinfo
  {author} {\bibfnamefont {N.}~\bibnamefont {Sun}}, \bibinfo {author}
  {\bibfnamefont {J.}~\bibnamefont {Wang}}, \bibinfo {author} {\bibfnamefont
  {C.}~\bibnamefont {Wu}}, \bibinfo {author} {\bibfnamefont {Y.}~\bibnamefont
  {Chen}},\ and\ \bibinfo {author} {\bibfnamefont {O.}~\bibnamefont {Temam}},\
  }\href {https://doi.org/10.1145/2654822.2541967} {\bibfield  {journal}
  {\bibinfo  {journal} {ACM SIGARCH Computer Architecture News}\ }\textbf
  {\bibinfo {volume} {42}},\ \bibinfo {pages} {269} (\bibinfo {year}
  {2014})}\BibitemShut {NoStop}%
\bibitem [{\citenamefont {Vivien}\ \emph {et~al.}(2012)\citenamefont {Vivien},
  \citenamefont {Polzer}, \citenamefont {Marris-Morini}, \citenamefont
  {Osmond}, \citenamefont {Hartmann}, \citenamefont {Crozat}, \citenamefont
  {Cassan}, \citenamefont {Kopp}, \citenamefont {Zimmermann},\ and\
  \citenamefont {Fédéli}}]{speed}%
  \BibitemOpen
  \bibfield  {author} {\bibinfo {author} {\bibfnamefont {L.}~\bibnamefont
  {Vivien}}, \bibinfo {author} {\bibfnamefont {A.}~\bibnamefont {Polzer}},
  \bibinfo {author} {\bibfnamefont {D.}~\bibnamefont {Marris-Morini}}, \bibinfo
  {author} {\bibfnamefont {J.}~\bibnamefont {Osmond}}, \bibinfo {author}
  {\bibfnamefont {J.~M.}\ \bibnamefont {Hartmann}}, \bibinfo {author}
  {\bibfnamefont {P.}~\bibnamefont {Crozat}}, \bibinfo {author} {\bibfnamefont
  {E.}~\bibnamefont {Cassan}}, \bibinfo {author} {\bibfnamefont
  {C.}~\bibnamefont {Kopp}}, \bibinfo {author} {\bibfnamefont {H.}~\bibnamefont
  {Zimmermann}},\ and\ \bibinfo {author} {\bibfnamefont {J.~M.}\ \bibnamefont
  {Fédéli}},\ }\href {https://doi.org/10.1364/OE.20.001096} {\bibfield
  {journal} {\bibinfo  {journal} {Opt. Express}\ }\textbf {\bibinfo {volume}
  {20}},\ \bibinfo {pages} {1096} (\bibinfo {year} {2012})}\BibitemShut
  {NoStop}%
\bibitem [{\citenamefont {Lin}\ \emph {et~al.}(2018)\citenamefont {Lin},
  \citenamefont {Rivenson}, \citenamefont {Yardimci}, \citenamefont {Veli},
  \citenamefont {Luo}, \citenamefont {Jarrahi},\ and\ \citenamefont
  {Ozcan}}]{ONN_histories_1}%
  \BibitemOpen
  \bibfield  {author} {\bibinfo {author} {\bibfnamefont {X.}~\bibnamefont
  {Lin}}, \bibinfo {author} {\bibfnamefont {Y.}~\bibnamefont {Rivenson}},
  \bibinfo {author} {\bibfnamefont {N.~T.}\ \bibnamefont {Yardimci}}, \bibinfo
  {author} {\bibfnamefont {M.}~\bibnamefont {Veli}}, \bibinfo {author}
  {\bibfnamefont {Y.}~\bibnamefont {Luo}}, \bibinfo {author} {\bibfnamefont
  {M.}~\bibnamefont {Jarrahi}},\ and\ \bibinfo {author} {\bibfnamefont
  {A.}~\bibnamefont {Ozcan}},\ }\href {https://doi.org/10.1126/science.aat8084}
  {\bibfield  {journal} {\bibinfo  {journal} {Science}\ }\textbf {\bibinfo
  {volume} {361}},\ \bibinfo {pages} {1004} (\bibinfo {year}
  {2018})}\BibitemShut {NoStop}%
\bibitem [{\citenamefont {Shen}\ \emph {et~al.}(2017)\citenamefont {Shen},
  \citenamefont {Harris}, \citenamefont {Skirlo}, \citenamefont {Prabhu},
  \citenamefont {Baehr-Jones}, \citenamefont {Hochberg}, \citenamefont {Sun},
  \citenamefont {Zhao}, \citenamefont {Larochelle},\ and\ \citenamefont
  {Englund}}]{ONN_histories_2}%
  \BibitemOpen
  \bibfield  {author} {\bibinfo {author} {\bibfnamefont {Y.}~\bibnamefont
  {Shen}}, \bibinfo {author} {\bibfnamefont {N.~C.}\ \bibnamefont {Harris}},
  \bibinfo {author} {\bibfnamefont {S.}~\bibnamefont {Skirlo}}, \bibinfo
  {author} {\bibfnamefont {M.}~\bibnamefont {Prabhu}}, \bibinfo {author}
  {\bibfnamefont {T.}~\bibnamefont {Baehr-Jones}}, \bibinfo {author}
  {\bibfnamefont {M.}~\bibnamefont {Hochberg}}, \bibinfo {author}
  {\bibfnamefont {X.}~\bibnamefont {Sun}}, \bibinfo {author} {\bibfnamefont
  {S.}~\bibnamefont {Zhao}}, \bibinfo {author} {\bibfnamefont {H.}~\bibnamefont
  {Larochelle}},\ and\ \bibinfo {author} {\bibfnamefont {D.}~\bibnamefont
  {Englund}},\ }\href {https://doi.org/10.1038/nphoton.2017.93} {\bibfield
  {journal} {\bibinfo  {journal} {Nature Photon.}\ }\textbf {\bibinfo {volume}
  {11}},\ \bibinfo {pages} {441} (\bibinfo {year} {2017})}\BibitemShut
  {NoStop}%
\bibitem [{\citenamefont {Tait}\ \emph {et~al.}(2017)\citenamefont {Tait},
  \citenamefont {Lima}, \citenamefont {X.}, \citenamefont {A.}, \citenamefont
  {J.},\ and\ \citenamefont {R.}}]{ONN_histories_3}%
  \BibitemOpen
  \bibfield  {author} {\bibinfo {author} {\bibfnamefont {A.~N.}\ \bibnamefont
  {Tait}}, \bibinfo {author} {\bibfnamefont {T.~F.}\ \bibnamefont {Lima}},
  \bibinfo {author} {\bibfnamefont {W.~A.}\ \bibnamefont {X.}}, \bibinfo
  {author} {\bibfnamefont {N.~M.}\ \bibnamefont {A.}}, \bibinfo {author}
  {\bibfnamefont {S.~B.}\ \bibnamefont {J.}},\ and\ \bibinfo {author}
  {\bibfnamefont {P.~P.}\ \bibnamefont {R.}},\ }\href
  {https://doi.org/10.1038/s41598-017-07754-z} {\bibfield  {journal} {\bibinfo
  {journal} {Sci Rep}\ }\textbf {\bibinfo {volume} {7}},\ \bibinfo {pages}
  {7430} (\bibinfo {year} {2017})}\BibitemShut {NoStop}%
\bibitem [{\citenamefont {Shi}\ \emph {et~al.}(2018)\citenamefont {Shi},
  \citenamefont {Calabretta}, \citenamefont {Bunandar}, \citenamefont
  {Englund},\ and\ \citenamefont {Stabile}}]{ONN_histories_4}%
  \BibitemOpen
  \bibfield  {author} {\bibinfo {author} {\bibfnamefont {B.}~\bibnamefont
  {Shi}}, \bibinfo {author} {\bibfnamefont {N.}~\bibnamefont {Calabretta}},
  \bibinfo {author} {\bibfnamefont {D.}~\bibnamefont {Bunandar}}, \bibinfo
  {author} {\bibfnamefont {D.}~\bibnamefont {Englund}},\ and\ \bibinfo {author}
  {\bibfnamefont {R.}~\bibnamefont {Stabile}},\ }in\ \href
  {https://doi.org/10.1109/PS.2018.8751338} {\emph {\bibinfo {booktitle} {2018
  Photonics in Switching and Computing (PSC)}}}\ (\bibinfo {year} {2018})\ pp.\
  \bibinfo {pages} {1--3}\BibitemShut {NoStop}%
\bibitem [{\citenamefont {Zuo}\ \emph {et~al.}(2019)\citenamefont {Zuo},
  \citenamefont {Li}, \citenamefont {Zhao}, \citenamefont {Chen}, \citenamefont
  {Jo}, \citenamefont {Liu},\ and\ \citenamefont {Du}}]{ONN_histories_5}%
  \BibitemOpen
  \bibfield  {author} {\bibinfo {author} {\bibfnamefont {Y.}~\bibnamefont
  {Zuo}}, \bibinfo {author} {\bibfnamefont {B.}~\bibnamefont {Li}}, \bibinfo
  {author} {\bibfnamefont {Y.}~\bibnamefont {Zhao}}, \bibinfo {author}
  {\bibfnamefont {Y.~C.}\ \bibnamefont {Chen}}, \bibinfo {author}
  {\bibfnamefont {G.~B.}\ \bibnamefont {Jo}}, \bibinfo {author} {\bibfnamefont
  {J.}~\bibnamefont {Liu}},\ and\ \bibinfo {author} {\bibfnamefont
  {S.}~\bibnamefont {Du}},\ }\href {https://doi.org/10.1364/OPTICA.6.001132}
  {\bibfield  {journal} {\bibinfo  {journal} {Optica}\ }\textbf {\bibinfo
  {volume} {6}},\ \bibinfo {pages} {1132} (\bibinfo {year} {2019})}\BibitemShut
  {NoStop}%
\bibitem [{\citenamefont {Borghi}\ \emph {et~al.}(2019)\citenamefont {Borghi},
  \citenamefont {Trenti},\ and\ \citenamefont {Pavesi}}]{FWM}%
  \BibitemOpen
  \bibfield  {author} {\bibinfo {author} {\bibfnamefont {M.}~\bibnamefont
  {Borghi}}, \bibinfo {author} {\bibfnamefont {A.}~\bibnamefont {Trenti}},\
  and\ \bibinfo {author} {\bibfnamefont {L.}~\bibnamefont {Pavesi}},\ }\href
  {https://doi.org/10.1038/s41598-018-36694-5} {\bibfield  {journal} {\bibinfo
  {journal} {Scientific reports}\ }\textbf {\bibinfo {volume} {9}},\ \bibinfo
  {pages} {1} (\bibinfo {year} {2019})}\BibitemShut {NoStop}%
\bibitem [{\citenamefont {Kibria}\ and\ \citenamefont
  {Austin}(2015)}]{FWM_app}%
  \BibitemOpen
  \bibfield  {author} {\bibinfo {author} {\bibfnamefont {R.}~\bibnamefont
  {Kibria}}\ and\ \bibinfo {author} {\bibfnamefont {M.}~\bibnamefont
  {Austin}},\ }\href {https://doi.org/10.3390/photonics2010200} {\bibfield
  {journal} {\bibinfo  {journal} {Photonics}\ }\textbf {\bibinfo {volume}
  {2}},\ \bibinfo {pages} {200} (\bibinfo {year} {2015})}\BibitemShut {NoStop}%
\bibitem [{\citenamefont {Mourgias-Alexandris}\ \emph
  {et~al.}(2019)\citenamefont {Mourgias-Alexandris}, \citenamefont
  {Tsakyridis}, \citenamefont {Passalis}, \citenamefont {Tefas}, \citenamefont
  {Vyrsokinos},\ and\ \citenamefont {Pleros}}]{NL_O_1}%
  \BibitemOpen
  \bibfield  {author} {\bibinfo {author} {\bibfnamefont {G.}~\bibnamefont
  {Mourgias-Alexandris}}, \bibinfo {author} {\bibfnamefont {A.}~\bibnamefont
  {Tsakyridis}}, \bibinfo {author} {\bibfnamefont {N.}~\bibnamefont
  {Passalis}}, \bibinfo {author} {\bibfnamefont {A.}~\bibnamefont {Tefas}},
  \bibinfo {author} {\bibfnamefont {K.}~\bibnamefont {Vyrsokinos}},\ and\
  \bibinfo {author} {\bibfnamefont {N.}~\bibnamefont {Pleros}},\ }\href
  {https://doi.org/10.1364/OE.27.009620} {\bibfield  {journal} {\bibinfo
  {journal} {Optics express}\ }\textbf {\bibinfo {volume} {27}},\ \bibinfo
  {pages} {9620} (\bibinfo {year} {2019})}\BibitemShut {NoStop}%
\bibitem [{\citenamefont {Hill}\ \emph {et~al.}(2002)\citenamefont {Hill},
  \citenamefont {Frietman}, \citenamefont {de~Waardt}, \citenamefont {Khoe},\
  and\ \citenamefont {Dorren}}]{NL_O_2}%
  \BibitemOpen
  \bibfield  {author} {\bibinfo {author} {\bibfnamefont {M.~T.}\ \bibnamefont
  {Hill}}, \bibinfo {author} {\bibfnamefont {E.~E.}\ \bibnamefont {Frietman}},
  \bibinfo {author} {\bibfnamefont {H.}~\bibnamefont {de~Waardt}}, \bibinfo
  {author} {\bibfnamefont {G.-d.}\ \bibnamefont {Khoe}},\ and\ \bibinfo
  {author} {\bibfnamefont {H.~J.}\ \bibnamefont {Dorren}},\ }\href
  {https://doi.org/10.1109/TNN.2002.804222} {\bibfield  {journal} {\bibinfo
  {journal} {IEEE Transactions on Neural Networks}\ }\textbf {\bibinfo {volume}
  {13}},\ \bibinfo {pages} {1504} (\bibinfo {year} {2002})}\BibitemShut
  {NoStop}%
\bibitem [{\citenamefont {Rosenbluth}\ \emph {et~al.}(2009)\citenamefont
  {Rosenbluth}, \citenamefont {Kravtsov}, \citenamefont {Fok},\ and\
  \citenamefont {Prucnal}}]{NL_O_3}%
  \BibitemOpen
  \bibfield  {author} {\bibinfo {author} {\bibfnamefont {D.}~\bibnamefont
  {Rosenbluth}}, \bibinfo {author} {\bibfnamefont {K.}~\bibnamefont
  {Kravtsov}}, \bibinfo {author} {\bibfnamefont {M.~P.}\ \bibnamefont {Fok}},\
  and\ \bibinfo {author} {\bibfnamefont {P.~R.}\ \bibnamefont {Prucnal}},\
  }\href {https://doi.org/10.1364/OE.17.022767} {\bibfield  {journal} {\bibinfo
   {journal} {Optics express}\ }\textbf {\bibinfo {volume} {17}},\ \bibinfo
  {pages} {22767} (\bibinfo {year} {2009})}\BibitemShut {NoStop}%
\bibitem [{\citenamefont {Miscuglio}\ \emph {et~al.}(2018)\citenamefont
  {Miscuglio}, \citenamefont {Mehrabian}, \citenamefont {Hu}, \citenamefont
  {Azzam}, \citenamefont {George}, \citenamefont {Kildishev}, \citenamefont
  {Pelton},\ and\ \citenamefont {Sorger}}]{NL_O_4}%
  \BibitemOpen
  \bibfield  {author} {\bibinfo {author} {\bibfnamefont {M.}~\bibnamefont
  {Miscuglio}}, \bibinfo {author} {\bibfnamefont {A.}~\bibnamefont
  {Mehrabian}}, \bibinfo {author} {\bibfnamefont {Z.}~\bibnamefont {Hu}},
  \bibinfo {author} {\bibfnamefont {S.~I.}\ \bibnamefont {Azzam}}, \bibinfo
  {author} {\bibfnamefont {J.}~\bibnamefont {George}}, \bibinfo {author}
  {\bibfnamefont {A.~V.}\ \bibnamefont {Kildishev}}, \bibinfo {author}
  {\bibfnamefont {M.}~\bibnamefont {Pelton}},\ and\ \bibinfo {author}
  {\bibfnamefont {V.~J.}\ \bibnamefont {Sorger}},\ }\href
  {https://doi.org/10.1364/OME.8.003851} {\bibfield  {journal} {\bibinfo
  {journal} {Optical Materials Express}\ }\textbf {\bibinfo {volume} {8}},\
  \bibinfo {pages} {3851} (\bibinfo {year} {2018})}\BibitemShut {NoStop}%
\bibitem [{\citenamefont {Tait}\ \emph {et~al.}(2019)\citenamefont {Tait},
  \citenamefont {De~Lima}, \citenamefont {Nahmias}, \citenamefont {Miller},
  \citenamefont {Peng}, \citenamefont {Shastri},\ and\ \citenamefont
  {Prucnal}}]{NL_OEO_1}%
  \BibitemOpen
  \bibfield  {author} {\bibinfo {author} {\bibfnamefont {A.~N.}\ \bibnamefont
  {Tait}}, \bibinfo {author} {\bibfnamefont {T.~F.}\ \bibnamefont {De~Lima}},
  \bibinfo {author} {\bibfnamefont {M.~A.}\ \bibnamefont {Nahmias}}, \bibinfo
  {author} {\bibfnamefont {H.~B.}\ \bibnamefont {Miller}}, \bibinfo {author}
  {\bibfnamefont {H.-T.}\ \bibnamefont {Peng}}, \bibinfo {author}
  {\bibfnamefont {B.~J.}\ \bibnamefont {Shastri}},\ and\ \bibinfo {author}
  {\bibfnamefont {P.~R.}\ \bibnamefont {Prucnal}},\ }\href
  {https://doi.org/10.1103/PhysRevApplied.11.064043} {\bibfield  {journal}
  {\bibinfo  {journal} {Physical Review Applied}\ }\textbf {\bibinfo {volume}
  {11}},\ \bibinfo {pages} {064043} (\bibinfo {year} {2019})}\BibitemShut
  {NoStop}%
\bibitem [{\citenamefont {Amin}\ \emph {et~al.}(2019)\citenamefont {Amin},
  \citenamefont {George}, \citenamefont {Sun}, \citenamefont {Ferreira~de
  Lima}, \citenamefont {Tait}, \citenamefont {Khurgin}, \citenamefont
  {Miscuglio}, \citenamefont {Shastri}, \citenamefont {Prucnal}, \citenamefont
  {El-Ghazawi} \emph {et~al.}}]{NL_OEO_2}%
  \BibitemOpen
  \bibfield  {author} {\bibinfo {author} {\bibfnamefont {R.}~\bibnamefont
  {Amin}}, \bibinfo {author} {\bibfnamefont {J.}~\bibnamefont {George}},
  \bibinfo {author} {\bibfnamefont {S.}~\bibnamefont {Sun}}, \bibinfo {author}
  {\bibfnamefont {T.}~\bibnamefont {Ferreira~de Lima}}, \bibinfo {author}
  {\bibfnamefont {A.~N.}\ \bibnamefont {Tait}}, \bibinfo {author}
  {\bibfnamefont {J.}~\bibnamefont {Khurgin}}, \bibinfo {author} {\bibfnamefont
  {M.}~\bibnamefont {Miscuglio}}, \bibinfo {author} {\bibfnamefont {B.~J.}\
  \bibnamefont {Shastri}}, \bibinfo {author} {\bibfnamefont {P.~R.}\
  \bibnamefont {Prucnal}}, \bibinfo {author} {\bibfnamefont {T.}~\bibnamefont
  {El-Ghazawi}}, \emph {et~al.},\ }\href {https://doi.org/10.1063/1.5109039}
  {\bibfield  {journal} {\bibinfo  {journal} {APL Materials}\ }\textbf
  {\bibinfo {volume} {7}},\ \bibinfo {pages} {081112} (\bibinfo {year}
  {2019})}\BibitemShut {NoStop}%
\bibitem [{\citenamefont {George}\ \emph {et~al.}(2019)\citenamefont {George},
  \citenamefont {Mehrabian}, \citenamefont {Amin}, \citenamefont {Meng},
  \citenamefont {De~Lima}, \citenamefont {Tait}, \citenamefont {Shastri},
  \citenamefont {El-Ghazawi}, \citenamefont {Prucnal},\ and\ \citenamefont
  {Sorger}}]{NL_OEO_3}%
  \BibitemOpen
  \bibfield  {author} {\bibinfo {author} {\bibfnamefont {J.~K.}\ \bibnamefont
  {George}}, \bibinfo {author} {\bibfnamefont {A.}~\bibnamefont {Mehrabian}},
  \bibinfo {author} {\bibfnamefont {R.}~\bibnamefont {Amin}}, \bibinfo {author}
  {\bibfnamefont {J.}~\bibnamefont {Meng}}, \bibinfo {author} {\bibfnamefont
  {T.~F.}\ \bibnamefont {De~Lima}}, \bibinfo {author} {\bibfnamefont {A.~N.}\
  \bibnamefont {Tait}}, \bibinfo {author} {\bibfnamefont {B.~J.}\ \bibnamefont
  {Shastri}}, \bibinfo {author} {\bibfnamefont {T.}~\bibnamefont {El-Ghazawi}},
  \bibinfo {author} {\bibfnamefont {P.~R.}\ \bibnamefont {Prucnal}},\ and\
  \bibinfo {author} {\bibfnamefont {V.~J.}\ \bibnamefont {Sorger}},\ }\href
  {https://doi.org/10.1063/1.5109039} {\bibfield  {journal} {\bibinfo
  {journal} {Optics express}\ }\textbf {\bibinfo {volume} {27}},\ \bibinfo
  {pages} {5181} (\bibinfo {year} {2019})}\BibitemShut {NoStop}%
\bibitem [{\citenamefont {Nahmias}\ \emph {et~al.}(2016)\citenamefont
  {Nahmias}, \citenamefont {Tait}, \citenamefont {Tolias}, \citenamefont
  {Chang}, \citenamefont {Ferreira~de Lima}, \citenamefont {Shastri},\ and\
  \citenamefont {Prucnal}}]{NL_OEO_4}%
  \BibitemOpen
  \bibfield  {author} {\bibinfo {author} {\bibfnamefont {M.~A.}\ \bibnamefont
  {Nahmias}}, \bibinfo {author} {\bibfnamefont {A.~N.}\ \bibnamefont {Tait}},
  \bibinfo {author} {\bibfnamefont {L.}~\bibnamefont {Tolias}}, \bibinfo
  {author} {\bibfnamefont {M.~P.}\ \bibnamefont {Chang}}, \bibinfo {author}
  {\bibfnamefont {T.}~\bibnamefont {Ferreira~de Lima}}, \bibinfo {author}
  {\bibfnamefont {B.~J.}\ \bibnamefont {Shastri}},\ and\ \bibinfo {author}
  {\bibfnamefont {P.~R.}\ \bibnamefont {Prucnal}},\ }\href
  {https://doi.org/10.1063/1.4945368} {\bibfield  {journal} {\bibinfo
  {journal} {Applied Physics Letters}\ }\textbf {\bibinfo {volume} {108}},\
  \bibinfo {pages} {151106} (\bibinfo {year} {2016})}\BibitemShut {NoStop}%
\bibitem [{\citenamefont {Williamson}\ \emph {et~al.}(2019)\citenamefont
  {Williamson}, \citenamefont {Hughes}, \citenamefont {Minkov}, \citenamefont
  {Bartlett}, \citenamefont {Pai},\ and\ \citenamefont {Fan}}]{NL_OEO_5}%
  \BibitemOpen
  \bibfield  {author} {\bibinfo {author} {\bibfnamefont {I.~A.}\ \bibnamefont
  {Williamson}}, \bibinfo {author} {\bibfnamefont {T.~W.}\ \bibnamefont
  {Hughes}}, \bibinfo {author} {\bibfnamefont {M.}~\bibnamefont {Minkov}},
  \bibinfo {author} {\bibfnamefont {B.}~\bibnamefont {Bartlett}}, \bibinfo
  {author} {\bibfnamefont {S.}~\bibnamefont {Pai}},\ and\ \bibinfo {author}
  {\bibfnamefont {S.}~\bibnamefont {Fan}},\ }\href
  {https://doi.org/10.1109/JSTQE.2019.2930455} {\bibfield  {journal} {\bibinfo
  {journal} {IEEE Journal of Selected Topics in Quantum Electronics}\ }\textbf
  {\bibinfo {volume} {26}},\ \bibinfo {pages} {1} (\bibinfo {year}
  {2019})}\BibitemShut {NoStop}%
\bibitem [{\citenamefont {Fisher}(1936)}]{Iris}%
  \BibitemOpen
  \bibfield  {author} {\bibinfo {author} {\bibfnamefont {R.~A.}\ \bibnamefont
  {Fisher}},\ }\href {https://doi.org/10.1111/j.1469-1809.1936.tb02137.x}
  {\bibfield  {journal} {\bibinfo  {journal} {Annals of eugenics}\ }\textbf
  {\bibinfo {volume} {7}},\ \bibinfo {pages} {179} (\bibinfo {year}
  {1936})}\BibitemShut {NoStop}%
\bibitem [{\citenamefont {Deng}(2012)}]{MNIST}%
  \BibitemOpen
  \bibfield  {author} {\bibinfo {author} {\bibfnamefont {L.}~\bibnamefont
  {Deng}},\ }\href {https://doi.org/10.1109/MSP.2012.2211477} {\bibfield
  {journal} {\bibinfo  {journal} {IEEE Signal Processing Magazine}\ }\textbf
  {\bibinfo {volume} {29}},\ \bibinfo {pages} {141} (\bibinfo {year}
  {2012})}\BibitemShut {NoStop}%
\bibitem [{\citenamefont {Heuck}\ \emph {et~al.}(2019)\citenamefont {Heuck},
  \citenamefont {Koefoed}, \citenamefont {Christensen}, \citenamefont {Ding},
  \citenamefont {Frandsen}, \citenamefont {Rottwitt},\ and\ \citenamefont
  {Oxenl{\o}we}}]{FWM_equations}%
  \BibitemOpen
  \bibfield  {author} {\bibinfo {author} {\bibfnamefont {M.}~\bibnamefont
  {Heuck}}, \bibinfo {author} {\bibfnamefont {J.~G.}\ \bibnamefont {Koefoed}},
  \bibinfo {author} {\bibfnamefont {J.~B.}\ \bibnamefont {Christensen}},
  \bibinfo {author} {\bibfnamefont {Y.}~\bibnamefont {Ding}}, \bibinfo {author}
  {\bibfnamefont {L.~H.}\ \bibnamefont {Frandsen}}, \bibinfo {author}
  {\bibfnamefont {K.}~\bibnamefont {Rottwitt}},\ and\ \bibinfo {author}
  {\bibfnamefont {L.~K.}\ \bibnamefont {Oxenl{\o}we}},\ }\href
  {https://doi.org/10.1088/1367-2630/ab09a7} {\bibfield  {journal} {\bibinfo
  {journal} {New Journal of Physics}\ }\textbf {\bibinfo {volume} {21}},\
  \bibinfo {pages} {033037} (\bibinfo {year} {2019})}\BibitemShut {NoStop}%
\bibitem [{\citenamefont {Suh}\ \emph {et~al.}(2004)\citenamefont {Suh},
  \citenamefont {Wang},\ and\ \citenamefont {Fan}}]{CMT}%
  \BibitemOpen
  \bibfield  {author} {\bibinfo {author} {\bibfnamefont {W.}~\bibnamefont
  {Suh}}, \bibinfo {author} {\bibfnamefont {Z.}~\bibnamefont {Wang}},\ and\
  \bibinfo {author} {\bibfnamefont {S.}~\bibnamefont {Fan}},\ }\href
  {https://doi.org/10.1109/JQE.2004.834773} {\bibfield  {journal} {\bibinfo
  {journal} {IEEE Journal of Quantum Electronics}\ }\textbf {\bibinfo {volume}
  {40}},\ \bibinfo {pages} {1511} (\bibinfo {year} {2004})}\BibitemShut
  {NoStop}%
\bibitem [{\citenamefont {Cappellini}\ and\ \citenamefont
  {Trillo}(1991)}]{TWM}%
  \BibitemOpen
  \bibfield  {author} {\bibinfo {author} {\bibfnamefont {G.}~\bibnamefont
  {Cappellini}}\ and\ \bibinfo {author} {\bibfnamefont {S.}~\bibnamefont
  {Trillo}},\ }\href {https://doi.org/10.1364/JOSAB.8.000824} {\bibfield
  {journal} {\bibinfo  {journal} {J. Opt. Soc. Am. B}\ }\textbf {\bibinfo
  {volume} {8}},\ \bibinfo {pages} {824} (\bibinfo {year} {1991})}\BibitemShut
  {NoStop}%
\bibitem [{\citenamefont {Zhang}\ \emph {et~al.}(2019)\citenamefont {Zhang},
  \citenamefont {Buscaino}, \citenamefont {Wang}, \citenamefont {Shams-Ansari},
  \citenamefont {Reimer}, \citenamefont {Zhu}, \citenamefont {Kahn},\ and\
  \citenamefont {Lon{\v{c}}ar}}]{EO_comb}%
  \BibitemOpen
  \bibfield  {author} {\bibinfo {author} {\bibfnamefont {M.}~\bibnamefont
  {Zhang}}, \bibinfo {author} {\bibfnamefont {B.}~\bibnamefont {Buscaino}},
  \bibinfo {author} {\bibfnamefont {C.}~\bibnamefont {Wang}}, \bibinfo {author}
  {\bibfnamefont {A.}~\bibnamefont {Shams-Ansari}}, \bibinfo {author}
  {\bibfnamefont {C.}~\bibnamefont {Reimer}}, \bibinfo {author} {\bibfnamefont
  {R.}~\bibnamefont {Zhu}}, \bibinfo {author} {\bibfnamefont {J.~M.}\
  \bibnamefont {Kahn}},\ and\ \bibinfo {author} {\bibfnamefont
  {M.}~\bibnamefont {Lon{\v{c}}ar}},\ }\href
  {https://doi.org/10.1038/s41586-019-1008-7} {\bibfield  {journal} {\bibinfo
  {journal} {Nature}\ }\textbf {\bibinfo {volume} {568}},\ \bibinfo {pages}
  {373} (\bibinfo {year} {2019})}\BibitemShut {NoStop}%
\bibitem [{\citenamefont {Kaini}\ \emph {et~al.}(2020)\citenamefont {Kaini},
  \citenamefont {Mbonde}, \citenamefont {Frankis}, \citenamefont {Mateman},
  \citenamefont {Leinse}, \citenamefont {Knights},\ and\ \citenamefont
  {Bradley}}]{FWM_ring_1}%
  \BibitemOpen
  \bibfield  {author} {\bibinfo {author} {\bibfnamefont {K.~M.}\ \bibnamefont
  {Kaini}}, \bibinfo {author} {\bibfnamefont {H.~M.}\ \bibnamefont {Mbonde}},
  \bibinfo {author} {\bibfnamefont {H.~C.}\ \bibnamefont {Frankis}}, \bibinfo
  {author} {\bibfnamefont {R.}~\bibnamefont {Mateman}}, \bibinfo {author}
  {\bibfnamefont {A.}~\bibnamefont {Leinse}}, \bibinfo {author} {\bibfnamefont
  {A.~P.}\ \bibnamefont {Knights}},\ and\ \bibinfo {author} {\bibfnamefont
  {J.~D.~B.}\ \bibnamefont {Bradley}},\ }\href
  {https://doi.org/10.1364/OSAC.402652} {\bibfield  {journal} {\bibinfo
  {journal} {OSA Continuum}\ }\textbf {\bibinfo {volume} {3}},\ \bibinfo
  {pages} {3497} (\bibinfo {year} {2020})}\BibitemShut {NoStop}%
\bibitem [{\citenamefont {Nicolás}\ and\ \citenamefont {Sipe}(2017)}]{q_nl}%
  \BibitemOpen
  \bibfield  {author} {\bibinfo {author} {\bibfnamefont {Q.}~\bibnamefont
  {Nicolás}}\ and\ \bibinfo {author} {\bibfnamefont {J.~E.}\ \bibnamefont
  {Sipe}},\ }\href {https://doi.org/10.1364/OL.42.003443} {\bibfield  {journal}
  {\bibinfo  {journal} {Opt. Lett.}\ }\textbf {\bibinfo {volume} {42}},\
  \bibinfo {pages} {3443} (\bibinfo {year} {2017})}\BibitemShut {NoStop}%
\bibitem [{\citenamefont {Paschotta}(2008)}]{rpphotonics}%
  \BibitemOpen
  \bibfield  {author} {\bibinfo {author} {\bibfnamefont {R.}~\bibnamefont
  {Paschotta}},\ }\href {https://www.rp-photonics.com/nonlinear_index.html}
  {\emph {\bibinfo {title} {Encyclopedia of laser physics and technology}}},\
  Vol.~\bibinfo {volume} {1}\ (\bibinfo  {publisher} {Wiley-vch},\ \bibinfo
  {year} {2008})\BibitemShut {NoStop}%
\bibitem [{\citenamefont {Ji}\ \emph {et~al.}(2017)\citenamefont {Ji},
  \citenamefont {Barbosa}, \citenamefont {Roberts}, \citenamefont {Dutt},
  \citenamefont {Cardenas}, \citenamefont {Okawachi}, \citenamefont {Bryant},
  \citenamefont {Gaeta},\ and\ \citenamefont {Lipson}}]{FWM_ring_2}%
  \BibitemOpen
  \bibfield  {author} {\bibinfo {author} {\bibfnamefont {X.}~\bibnamefont
  {Ji}}, \bibinfo {author} {\bibfnamefont {F.~A.}\ \bibnamefont {Barbosa}},
  \bibinfo {author} {\bibfnamefont {S.~P.}\ \bibnamefont {Roberts}}, \bibinfo
  {author} {\bibfnamefont {A.}~\bibnamefont {Dutt}}, \bibinfo {author}
  {\bibfnamefont {J.}~\bibnamefont {Cardenas}}, \bibinfo {author}
  {\bibfnamefont {Y.}~\bibnamefont {Okawachi}}, \bibinfo {author}
  {\bibfnamefont {A.}~\bibnamefont {Bryant}}, \bibinfo {author} {\bibfnamefont
  {A.~L.}\ \bibnamefont {Gaeta}},\ and\ \bibinfo {author} {\bibfnamefont
  {M.}~\bibnamefont {Lipson}},\ }\href
  {https://doi.org/10.1364/OPTICA.4.000619} {\bibfield  {journal} {\bibinfo
  {journal} {Optica}\ }\textbf {\bibinfo {volume} {4}},\ \bibinfo {pages} {619}
  (\bibinfo {year} {2017})}\BibitemShut {NoStop}%
\bibitem [{\citenamefont {Pour~Fard}\ \emph {et~al.}(2020)\citenamefont
  {Pour~Fard}, \citenamefont {Williamson}, \citenamefont {Edwards},
  \citenamefont {Liu}, \citenamefont {Pai}, \citenamefont {Bartlett},
  \citenamefont {Minkov}, \citenamefont {Hughes}, \citenamefont {Fan},\ and\
  \citenamefont {Nguyen}}]{thermo_optic}%
  \BibitemOpen
  \bibfield  {author} {\bibinfo {author} {\bibfnamefont {M.~M.}\ \bibnamefont
  {Pour~Fard}}, \bibinfo {author} {\bibfnamefont {I.~A.~D.}\ \bibnamefont
  {Williamson}}, \bibinfo {author} {\bibfnamefont {M.}~\bibnamefont {Edwards}},
  \bibinfo {author} {\bibfnamefont {K.}~\bibnamefont {Liu}}, \bibinfo {author}
  {\bibfnamefont {S.}~\bibnamefont {Pai}}, \bibinfo {author} {\bibfnamefont
  {B.}~\bibnamefont {Bartlett}}, \bibinfo {author} {\bibfnamefont
  {M.}~\bibnamefont {Minkov}}, \bibinfo {author} {\bibfnamefont {T.~W.}\
  \bibnamefont {Hughes}}, \bibinfo {author} {\bibfnamefont {S.}~\bibnamefont
  {Fan}},\ and\ \bibinfo {author} {\bibfnamefont {T.~A.}\ \bibnamefont
  {Nguyen}},\ }\href {https://doi.org/10.1364/OE.391473} {\bibfield  {journal}
  {\bibinfo  {journal} {Opt. Express}\ }\textbf {\bibinfo {volume} {28}},\
  \bibinfo {pages} {12138} (\bibinfo {year} {2020})}\BibitemShut {NoStop}%
\bibitem [{\citenamefont {George}\ \emph {et~al.}(2018)\citenamefont {George},
  \citenamefont {Amin}, \citenamefont {Mehrabian}, \citenamefont {Khurgin},
  \citenamefont {El-Ghazawi}, \citenamefont {Prucnal},\ and\ \citenamefont
  {Sorger}}]{nonlinear_acs_1}%
  \BibitemOpen
  \bibfield  {author} {\bibinfo {author} {\bibfnamefont {J.}~\bibnamefont
  {George}}, \bibinfo {author} {\bibfnamefont {R.}~\bibnamefont {Amin}},
  \bibinfo {author} {\bibfnamefont {A.}~\bibnamefont {Mehrabian}}, \bibinfo
  {author} {\bibfnamefont {J.}~\bibnamefont {Khurgin}}, \bibinfo {author}
  {\bibfnamefont {T.}~\bibnamefont {El-Ghazawi}}, \bibinfo {author}
  {\bibfnamefont {P.~R.}\ \bibnamefont {Prucnal}},\ and\ \bibinfo {author}
  {\bibfnamefont {V.~J.}\ \bibnamefont {Sorger}},\ }in\ \href
  {https://doi.org/10.1364/SPPCOM.2018.SpW4G.3} {\emph {\bibinfo {booktitle}
  {Signal Processing in Photonic Communications}}}\ (\bibinfo {organization}
  {Optical Society of America},\ \bibinfo {year} {2018})\ pp.\ \bibinfo {pages}
  {SpW4G--3}\BibitemShut {NoStop}%
\bibitem [{\citenamefont {George}\ \emph {et~al.}(2020)\citenamefont {George},
  \citenamefont {Amin}, \citenamefont {Meng}, \citenamefont {de~Lima},
  \citenamefont {Tait}, \citenamefont {Shastri}, \citenamefont {El-Ghazawi},
  \citenamefont {Prucnal},\ and\ \citenamefont {Sorger}}]{nonlinear_acs_2}%
  \BibitemOpen
  \bibfield  {author} {\bibinfo {author} {\bibfnamefont {J.~K.}\ \bibnamefont
  {George}}, \bibinfo {author} {\bibfnamefont {R.}~\bibnamefont {Amin}},
  \bibinfo {author} {\bibfnamefont {J.}~\bibnamefont {Meng}}, \bibinfo {author}
  {\bibfnamefont {T.~F.}\ \bibnamefont {de~Lima}}, \bibinfo {author}
  {\bibfnamefont {A.~N.}\ \bibnamefont {Tait}}, \bibinfo {author}
  {\bibfnamefont {B.~J.}\ \bibnamefont {Shastri}}, \bibinfo {author}
  {\bibfnamefont {T.}~\bibnamefont {El-Ghazawi}}, \bibinfo {author}
  {\bibfnamefont {P.~R.}\ \bibnamefont {Prucnal}},\ and\ \bibinfo {author}
  {\bibfnamefont {V.~J.}\ \bibnamefont {Sorger}},\ }\href
  {https://doi.org/10.1364/OE.391473} {\bibfield  {journal} {\bibinfo
  {journal} {Opt. Express}\ }\textbf {\bibinfo {volume} {28}},\ \bibinfo
  {pages} {12138} (\bibinfo {year} {2020})}\BibitemShut {NoStop}%
\bibitem [{\citenamefont {Rasmussen}\ \emph {et~al.}(2020)\citenamefont
  {Rasmussen}, \citenamefont {Yu},\ and\ \citenamefont
  {Mork}}]{nonlinear_acs_5}%
  \BibitemOpen
  \bibfield  {author} {\bibinfo {author} {\bibfnamefont {T.~S.}\ \bibnamefont
  {Rasmussen}}, \bibinfo {author} {\bibfnamefont {Y.}~\bibnamefont {Yu}},\ and\
  \bibinfo {author} {\bibfnamefont {J.}~\bibnamefont {Mork}},\ }\href
  {https://doi.org/10.1364/OL.395235} {\bibfield  {journal} {\bibinfo
  {journal} {Optics Letters}\ }\textbf {\bibinfo {volume} {45}},\ \bibinfo
  {pages} {3844} (\bibinfo {year} {2020})}\BibitemShut {NoStop}%
\bibitem [{\citenamefont {Mos}\ \emph {et~al.}(1997)\citenamefont {Mos},
  \citenamefont {Schleipen},\ and\ \citenamefont
  {De~Waardt}}]{nonlinear_acs_6}%
  \BibitemOpen
  \bibfield  {author} {\bibinfo {author} {\bibfnamefont {E.}~\bibnamefont
  {Mos}}, \bibinfo {author} {\bibfnamefont {J.}~\bibnamefont {Schleipen}},\
  and\ \bibinfo {author} {\bibfnamefont {H.}~\bibnamefont {De~Waardt}},\ }\href
  {https://doi.org/10.1364/AO.36.006654} {\bibfield  {journal} {\bibinfo
  {journal} {Applied optics}\ }\textbf {\bibinfo {volume} {36}},\ \bibinfo
  {pages} {6654} (\bibinfo {year} {1997})}\BibitemShut {NoStop}%
\bibitem [{\citenamefont {Dejonckheere}\ \emph {et~al.}(2014)\citenamefont
  {Dejonckheere}, \citenamefont {Duport}, \citenamefont {Smerieri},
  \citenamefont {Fang}, \citenamefont {Oudar}, \citenamefont {Haelterman},\
  and\ \citenamefont {Massar}}]{nonlinear_acs_3}%
  \BibitemOpen
  \bibfield  {author} {\bibinfo {author} {\bibfnamefont {A.}~\bibnamefont
  {Dejonckheere}}, \bibinfo {author} {\bibfnamefont {F.}~\bibnamefont
  {Duport}}, \bibinfo {author} {\bibfnamefont {A.}~\bibnamefont {Smerieri}},
  \bibinfo {author} {\bibfnamefont {L.}~\bibnamefont {Fang}}, \bibinfo {author}
  {\bibfnamefont {J.~L.}\ \bibnamefont {Oudar}}, \bibinfo {author}
  {\bibfnamefont {J.}~\bibnamefont {Haelterman}},\ and\ \bibinfo {author}
  {\bibfnamefont {S.}~\bibnamefont {Massar}},\ }\href
  {https://doi.org/10.1364/OE.22.010868} {\bibfield  {journal} {\bibinfo
  {journal} {Opt. Express}\ }\textbf {\bibinfo {volume} {22}},\ \bibinfo
  {pages} {10868} (\bibinfo {year} {2014})}\BibitemShut {NoStop}%
\bibitem [{\citenamefont {{Cheng}}\ \emph {et~al.}(2014)\citenamefont
  {{Cheng}}, \citenamefont {{Tsang}}, \citenamefont {{Wang}}, \citenamefont
  {{Xu}},\ and\ \citenamefont {{Xu}}}]{nonlinear_acs_4}%
  \BibitemOpen
  \bibfield  {author} {\bibinfo {author} {\bibfnamefont {Z.}~\bibnamefont
  {{Cheng}}}, \bibinfo {author} {\bibfnamefont {H.~K.}\ \bibnamefont
  {{Tsang}}}, \bibinfo {author} {\bibfnamefont {X.}~\bibnamefont {{Wang}}},
  \bibinfo {author} {\bibfnamefont {K.}~\bibnamefont {{Xu}}},\ and\ \bibinfo
  {author} {\bibfnamefont {J.}~\bibnamefont {{Xu}}},\ }\href
  {https://doi.org/10.1109/JSTQE.2013.2263115} {\bibfield  {journal} {\bibinfo
  {journal} {IEEE Journal of Selected Topics in Quantum Electronics}\ }\textbf
  {\bibinfo {volume} {20}},\ \bibinfo {pages} {43} (\bibinfo {year}
  {2014})}\BibitemShut {NoStop}%
\bibitem [{\citenamefont {Upham}\ \emph {et~al.}(2011)\citenamefont {Upham},
  \citenamefont {Tanaka}, \citenamefont {Kawamoto}, \citenamefont {Nakamura},
  \citenamefont {Song}, \citenamefont {Asano},\ and\ \citenamefont
  {Noda}}]{catch_release_1}%
  \BibitemOpen
  \bibfield  {author} {\bibinfo {author} {\bibfnamefont {J.}~\bibnamefont
  {Upham}}, \bibinfo {author} {\bibfnamefont {Y.}~\bibnamefont {Tanaka}},
  \bibinfo {author} {\bibfnamefont {Y.}~\bibnamefont {Kawamoto}, \bibfnamefont
  {Y.~Sato}}, \bibinfo {author} {\bibfnamefont {T.}~\bibnamefont {Nakamura}},
  \bibinfo {author} {\bibfnamefont {B.~S.}\ \bibnamefont {Song}}, \bibinfo
  {author} {\bibfnamefont {T.}~\bibnamefont {Asano}},\ and\ \bibinfo {author}
  {\bibfnamefont {S.}~\bibnamefont {Noda}},\ }\href
  {https://doi.org/10.1364/OE.19.023377} {\bibfield  {journal} {\bibinfo
  {journal} {Opt. Express}\ }\textbf {\bibinfo {volume} {19}},\ \bibinfo
  {pages} {23377} (\bibinfo {year} {2011})}\BibitemShut {NoStop}%
\bibitem [{\citenamefont {Nurdin}\ \emph {et~al.}(2016)\citenamefont {Nurdin},
  \citenamefont {James},\ and\ \citenamefont {Yamamoto}}]{catch_release_4}%
  \BibitemOpen
  \bibfield  {author} {\bibinfo {author} {\bibfnamefont {H.~I.}\ \bibnamefont
  {Nurdin}}, \bibinfo {author} {\bibfnamefont {M.~R.}\ \bibnamefont {James}},\
  and\ \bibinfo {author} {\bibfnamefont {N.}~\bibnamefont {Yamamoto}},\
  }\bibfield  {journal} {\bibinfo  {journal} {arXiv preprint}\ }\href
  {https://doi.org/arXiv:1609.05643} {arXiv:1609.05643} (\bibinfo {year}
  {2016})\BibitemShut {NoStop}%
\bibitem [{\citenamefont {Shaw}(1995)}]{PDE}%
  \BibitemOpen
  \bibfield  {author} {\bibinfo {author} {\bibfnamefont {K.~D.}\ \bibnamefont
  {Shaw}},\ }in\ \href {https://doi.org/10.1117/12.206492} {\emph {\bibinfo
  {booktitle} {Solid State Lasers and Nonlinear Crystals}}},\ Vol.\ \bibinfo
  {volume} {2379}\ (\bibinfo {organization} {SPIE},\ \bibinfo {year} {1995})\
  pp.\ \bibinfo {pages} {365--377}\BibitemShut {NoStop}%
\bibitem [{\citenamefont {Heuck}\ \emph {et~al.}(2003)\citenamefont {Heuck},
  \citenamefont {Kristensen}, \citenamefont {Elesin},\ and\ \citenamefont
  {Mørk}}]{catch_release_5}%
  \BibitemOpen
  \bibfield  {author} {\bibinfo {author} {\bibfnamefont {M.}~\bibnamefont
  {Heuck}}, \bibinfo {author} {\bibfnamefont {P.~T.}\ \bibnamefont
  {Kristensen}}, \bibinfo {author} {\bibfnamefont {Y.}~\bibnamefont {Elesin}},\
  and\ \bibinfo {author} {\bibfnamefont {J.}~\bibnamefont {Mørk}},\ }\href
  {https://doi.org/10.1364/OL.38.002466} {\bibfield  {journal} {\bibinfo
  {journal} {Opt. Lett.}\ }\textbf {\bibinfo {volume} {38}},\ \bibinfo {pages}
  {2466} (\bibinfo {year} {2003})}\BibitemShut {NoStop}%
\bibitem [{\citenamefont {Kristensen}\ \emph {et~al.}(2017)\citenamefont
  {Kristensen}, \citenamefont {de~Lasson}, \citenamefont {Heuck}, \citenamefont
  {Gregersen},\ and\ \citenamefont {Mørk}}]{catch_release_6}%
  \BibitemOpen
  \bibfield  {author} {\bibinfo {author} {\bibfnamefont {P.~T.}\ \bibnamefont
  {Kristensen}}, \bibinfo {author} {\bibfnamefont {J.~R.}\ \bibnamefont
  {de~Lasson}}, \bibinfo {author} {\bibfnamefont {M.}~\bibnamefont {Heuck}},
  \bibinfo {author} {\bibfnamefont {N.}~\bibnamefont {Gregersen}},\ and\
  \bibinfo {author} {\bibfnamefont {J.}~\bibnamefont {Mørk}},\ }\href
  {https://doi.org/10.1109/JLT.2017.2714263} {\bibfield  {journal} {\bibinfo
  {journal} {Journal of Lightwave Technology}\ }\textbf {\bibinfo {volume}
  {35}},\ \bibinfo {pages} {4247} (\bibinfo {year} {2017})}\BibitemShut
  {NoStop}%
\bibitem [{\citenamefont {Bergfeld}\ and\ \citenamefont {Daum}(2003)}]{GaAs}%
  \BibitemOpen
  \bibfield  {author} {\bibinfo {author} {\bibfnamefont {S.}~\bibnamefont
  {Bergfeld}}\ and\ \bibinfo {author} {\bibfnamefont {W.}~\bibnamefont
  {Daum}},\ }\href {https://doi.org/10.1103/PhysRevLett.90.036801} {\bibfield
  {journal} {\bibinfo  {journal} {Physical review letters}\ }\textbf {\bibinfo
  {volume} {90}},\ \bibinfo {pages} {036801} (\bibinfo {year}
  {2003})}\BibitemShut {NoStop}%
\bibitem [{\citenamefont {Yan}\ \emph {et~al.}(2022)\citenamefont {Yan},
  \citenamefont {He}, \citenamefont {Liu}, \citenamefont {Iu}, \citenamefont
  {Ahmed}, \citenamefont {Chen}, \citenamefont {Blakey}, \citenamefont
  {Akasaka}, \citenamefont {Ikeuchi},\ and\ \citenamefont {Helmy}}]{AlGaAs}%
  \BibitemOpen
  \bibfield  {author} {\bibinfo {author} {\bibfnamefont {Z.}~\bibnamefont
  {Yan}}, \bibinfo {author} {\bibfnamefont {H.}~\bibnamefont {He}}, \bibinfo
  {author} {\bibfnamefont {H.}~\bibnamefont {Liu}}, \bibinfo {author}
  {\bibfnamefont {M.}~\bibnamefont {Iu}}, \bibinfo {author} {\bibfnamefont
  {O.}~\bibnamefont {Ahmed}}, \bibinfo {author} {\bibfnamefont
  {E.}~\bibnamefont {Chen}}, \bibinfo {author} {\bibfnamefont {P.}~\bibnamefont
  {Blakey}}, \bibinfo {author} {\bibfnamefont {Y.}~\bibnamefont {Akasaka}},
  \bibinfo {author} {\bibfnamefont {T.}~\bibnamefont {Ikeuchi}},\ and\ \bibinfo
  {author} {\bibfnamefont {A.~S.}\ \bibnamefont {Helmy}},\ }\href
  {https://doi.org/10.1364/OPTICA.446674} {\bibfield  {journal} {\bibinfo
  {journal} {Optica}\ }\textbf {\bibinfo {volume} {9}},\ \bibinfo {pages} {56}
  (\bibinfo {year} {2022})}\BibitemShut {NoStop}%
\bibitem [{\citenamefont {Sato}\ \emph {et~al.}(2009)\citenamefont {Sato},
  \citenamefont {Abe}, \citenamefont {Shoji}, \citenamefont {Suda},\ and\
  \citenamefont {Kondo}}]{SiC}%
  \BibitemOpen
  \bibfield  {author} {\bibinfo {author} {\bibfnamefont {H.}~\bibnamefont
  {Sato}}, \bibinfo {author} {\bibfnamefont {M.}~\bibnamefont {Abe}}, \bibinfo
  {author} {\bibfnamefont {I.}~\bibnamefont {Shoji}}, \bibinfo {author}
  {\bibfnamefont {J.}~\bibnamefont {Suda}},\ and\ \bibinfo {author}
  {\bibfnamefont {T.}~\bibnamefont {Kondo}},\ }\href
  {https://doi.org/10.1364/JOSAB.26.001892} {\bibfield  {journal} {\bibinfo
  {journal} {JOSA B}\ }\textbf {\bibinfo {volume} {26}},\ \bibinfo {pages}
  {1892} (\bibinfo {year} {2009})}\BibitemShut {NoStop}%
\bibitem [{\citenamefont {Liu}\ \emph {et~al.}(2021)\citenamefont {Liu},
  \citenamefont {Huang}, \citenamefont {Wang}, \citenamefont {He},
  \citenamefont {Raja}, \citenamefont {Liu}, \citenamefont {Engelsen},\ and\
  \citenamefont {Kippenberg}}]{Single_platform_1}%
  \BibitemOpen
  \bibfield  {author} {\bibinfo {author} {\bibfnamefont {J.}~\bibnamefont
  {Liu}}, \bibinfo {author} {\bibfnamefont {G.}~\bibnamefont {Huang}}, \bibinfo
  {author} {\bibfnamefont {R.~N.}\ \bibnamefont {Wang}}, \bibinfo {author}
  {\bibfnamefont {J.}~\bibnamefont {He}}, \bibinfo {author} {\bibfnamefont
  {A.~S.}\ \bibnamefont {Raja}}, \bibinfo {author} {\bibfnamefont
  {T.}~\bibnamefont {Liu}}, \bibinfo {author} {\bibfnamefont {N.~J.}\
  \bibnamefont {Engelsen}},\ and\ \bibinfo {author} {\bibfnamefont {T.~J.}\
  \bibnamefont {Kippenberg}},\ }\href
  {https://doi.org/10.1038/s41467-021-21973-z} {\bibfield  {journal} {\bibinfo
  {journal} {Nature communications}\ }\textbf {\bibinfo {volume} {12}},\
  \bibinfo {pages} {1} (\bibinfo {year} {2021})}\BibitemShut {NoStop}%
\bibitem [{\citenamefont {Chang}\ \emph {et~al.}(2019)\citenamefont {Chang},
  \citenamefont {Boes}, \citenamefont {Pintus}, \citenamefont {Peters},
  \citenamefont {Kennedy}, \citenamefont {Jin}, \citenamefont {Guo},
  \citenamefont {Yu}, \citenamefont {Papp},\ and\ \citenamefont
  {Bowers}}]{Single_platform_2}%
  \BibitemOpen
  \bibfield  {author} {\bibinfo {author} {\bibfnamefont {L.}~\bibnamefont
  {Chang}}, \bibinfo {author} {\bibfnamefont {A.}~\bibnamefont {Boes}},
  \bibinfo {author} {\bibfnamefont {P.}~\bibnamefont {Pintus}}, \bibinfo
  {author} {\bibfnamefont {J.~D.}\ \bibnamefont {Peters}}, \bibinfo {author}
  {\bibfnamefont {M.}~\bibnamefont {Kennedy}}, \bibinfo {author} {\bibfnamefont
  {W.}~\bibnamefont {Jin}}, \bibinfo {author} {\bibfnamefont {X.-W.}\
  \bibnamefont {Guo}}, \bibinfo {author} {\bibfnamefont {S.-P.}\ \bibnamefont
  {Yu}}, \bibinfo {author} {\bibfnamefont {S.~B.}\ \bibnamefont {Papp}},\ and\
  \bibinfo {author} {\bibfnamefont {J.~E.}\ \bibnamefont {Bowers}},\ }in\ \href
  {https://doi.org/10.1364/CLEO_SI.2019.SF2I.2} {\emph {\bibinfo {booktitle}
  {CLEO: Science and Innovations}}}\ (\bibinfo {organization} {Optical Society
  of America},\ \bibinfo {year} {2019})\ pp.\ \bibinfo {pages}
  {SF2I--2}\BibitemShut {NoStop}%
\bibitem [{\citenamefont {Kingma}\ and\ \citenamefont {Ba}(2014)}]{Adam}%
  \BibitemOpen
  \bibfield  {author} {\bibinfo {author} {\bibfnamefont {D.~P.}\ \bibnamefont
  {Kingma}}\ and\ \bibinfo {author} {\bibfnamefont {J.}~\bibnamefont {Ba}},\
  }\bibfield  {journal} {\bibinfo  {journal} {arXiv preprint}\ }\href
  {https://doi.org/arXiv.1412.6980} {arXiv.1412.6980} (\bibinfo {year}
  {2014})\BibitemShut {NoStop}%
\bibitem [{\citenamefont {Guo}\ \emph {et~al.}(2021)\citenamefont {Guo},
  \citenamefont {Barrett}, \citenamefont {Wang},\ and\ \citenamefont
  {Lvovsky}}]{backprop}%
  \BibitemOpen
  \bibfield  {author} {\bibinfo {author} {\bibfnamefont {X.}~\bibnamefont
  {Guo}}, \bibinfo {author} {\bibfnamefont {T.~D.}\ \bibnamefont {Barrett}},
  \bibinfo {author} {\bibfnamefont {Z.~M.}\ \bibnamefont {Wang}},\ and\
  \bibinfo {author} {\bibfnamefont {A.}~\bibnamefont {Lvovsky}},\ }\href
  {https://doi.org/10.1364/PRJ.411104} {\bibfield  {journal} {\bibinfo
  {journal} {Photonics Research}\ }\textbf {\bibinfo {volume} {9}},\ \bibinfo
  {pages} {B71} (\bibinfo {year} {2021})}\BibitemShut {NoStop}%
\end{thebibliography}%

\clearpage

\begin{widetext}
\section{Appendix}
\end{widetext}

\subsection{Expressivity of the Realized Transformations}

In sections 2A and 2B, we studied the expressivity of the linear transformations realized in our proposed hardware. We do this by randomly sampling $M$ unitary matrices, each denoted by $U_{i}$ and for each one of them attempting to realize it in our hardware. The average overlap between target and realization, a.k.a. fidelity, a.k.a. expressivity is defined by Eq.~\eqref{eq:fid}. Here we present more detailed statistics over the sample of $M = 1000$ matrices, by giving a histogram of the single sample (single $U_i$) fidelities, instead of just their averages seen in Fig.~\ref{Passive_Express} and Fig.~\ref{Active_Express}.

\begin{figure}
\centering
\includegraphics[width = \columnwidth]{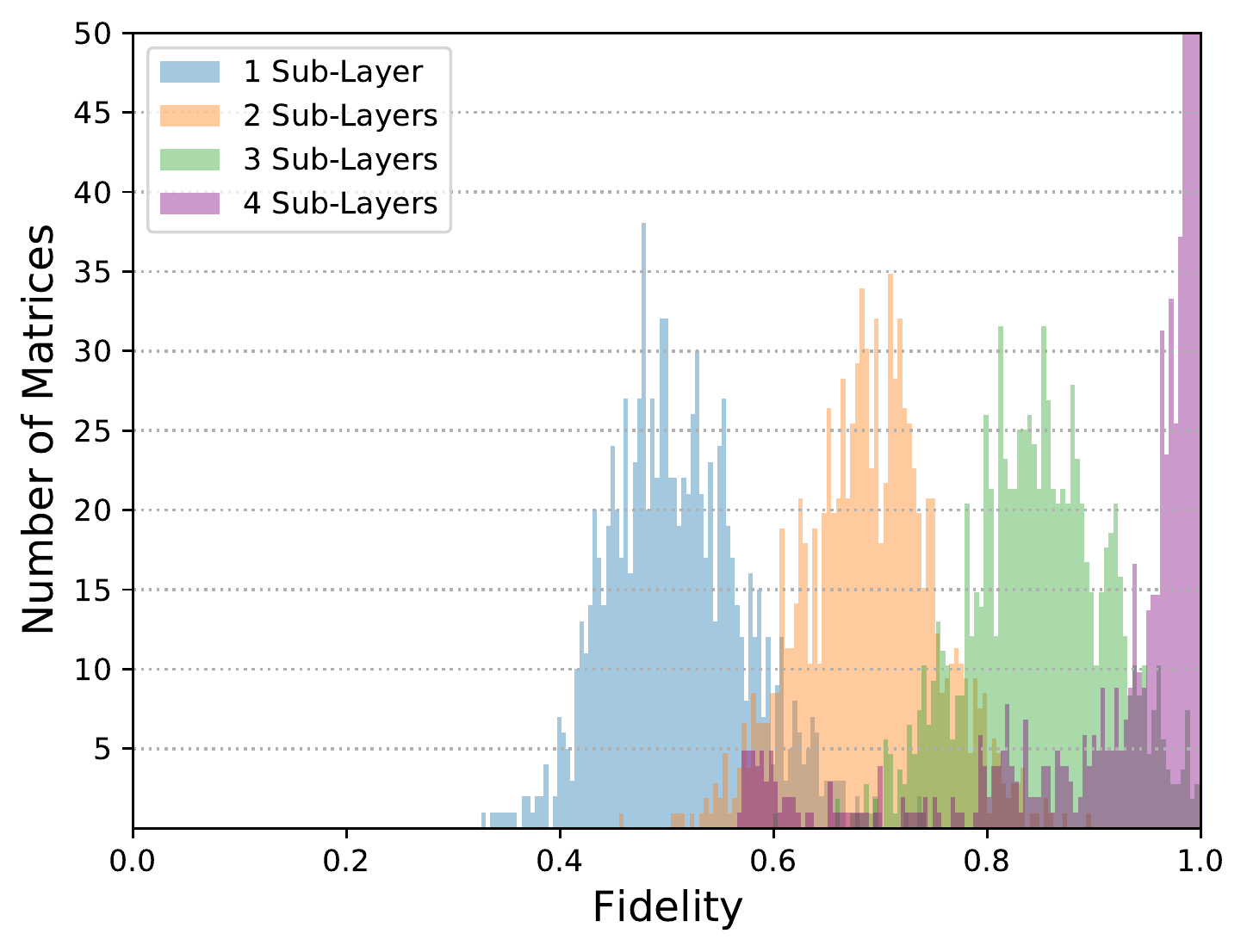}
\caption{\label{fid_passive_dist} Distribution of the fidelity of the optimized "passive coupling" transformations depending on the number of sub-layers utilized.}
\end{figure}

\begin{figure}
\centering
\includegraphics[width = \columnwidth]{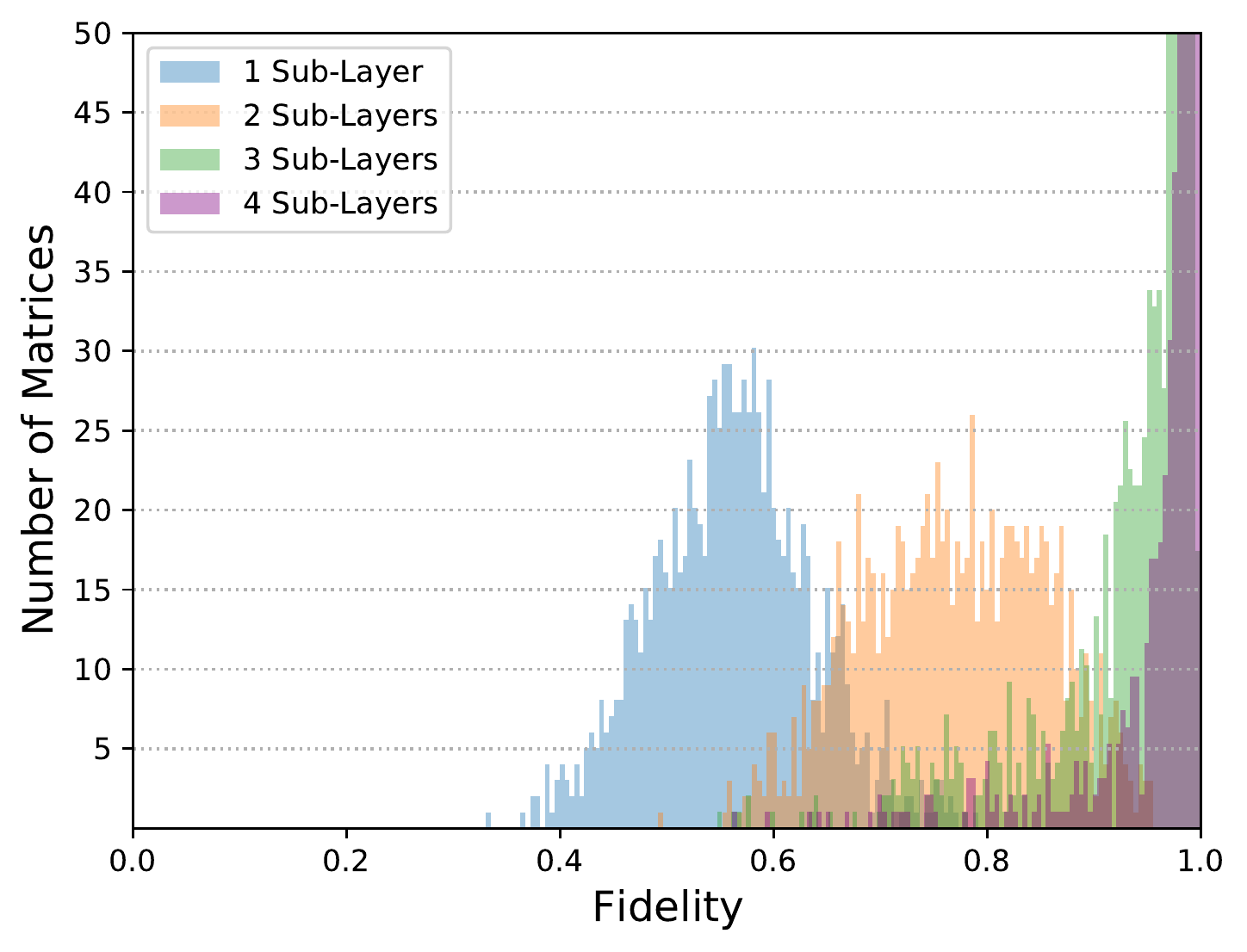}
\caption{\label{fid_active_dist} Distribution of the fidelity of the optimized "active coupling" transformations depending on the number of sub-layers utilized.}
\end{figure}

For the case of passive coupling, transformations are given by matrices of the form $\prod \textbf{T}$ from Eq.~\eqref{toeplitz}. Fig.~\ref{fid_passive_dist} shows the distribution of the number of matrices with the average fidelities increasing as the number of sub-layers are increased. We specifically chose small 4-by-4 matrices, as the behavior is easier to depict at that scale. Similar results can be seen for transformations of the type $\prod e^{\Delta t \textbf{P}}$ realised by the active coupling method, as seen in Fig.~\ref{fid_active_dist}.

\subsection{Deterministic Active Capture of a Pulse into a Resonator}

The dynamics of a incoming pulse $S_\mathrm{in}$ being captured into a resonator mode $A$ is described by Eqs.~\eqref{capture_1} and \eqref{capture_2}, where $S_\mathrm{out}$ is the outgoing pulse envelope. To ensure the entirety of the pulse is captured, we can rearrange the equations as
\begin{equation}
    S_{\mathrm{out}} = 0 \implies \sqrt{\gamma(t)} = \frac{S_{\mathrm{in}}}{A},
\end{equation}
\begin{equation}
    \frac{\mathrm{d}A}{\mathrm{d}t} = -\frac{\gamma_\mathrm{H}}{2}A + \frac{S_{i}^{2}}{2A}.
\end{equation}

We can solve this differential equation for an arbitrary incoming envelope. The analytical solution for an incoming pulse with a Gaussian envelope $S_\mathrm{in} = S_0e^{-\frac{(t - t_{\mathrm{0}})^{2}}{2w^{2}}}$ is $\sqrt{\gamma(t)} = \frac{S_0e^{-\frac{(t - t_{\mathrm{0}})^{2}}{2w^{2}}}}{A}$ where
\begin{multline}
    A(t) = \left(  \sqrt{\pi} S_{\mathrm{0}}w e^{\frac{\gamma_\mathrm{H}}{2}\left( \frac{\gamma_\mathrm{H}w^{2}}{2} + 2t_{\mathrm{0}} - 2t) \right)} \times \right. \\ \left. \frac{\left(1 + \mathrm{erf}\left(-\frac{\gamma_\mathrm{H}w^{2}}{2} - t_{0} + t\right) \right)}{2} \right)^{\frac{1}{2}}
    \label{analytic_sol}
\end{multline}

\begin{figure}
\centering
\includegraphics[width = \columnwidth]{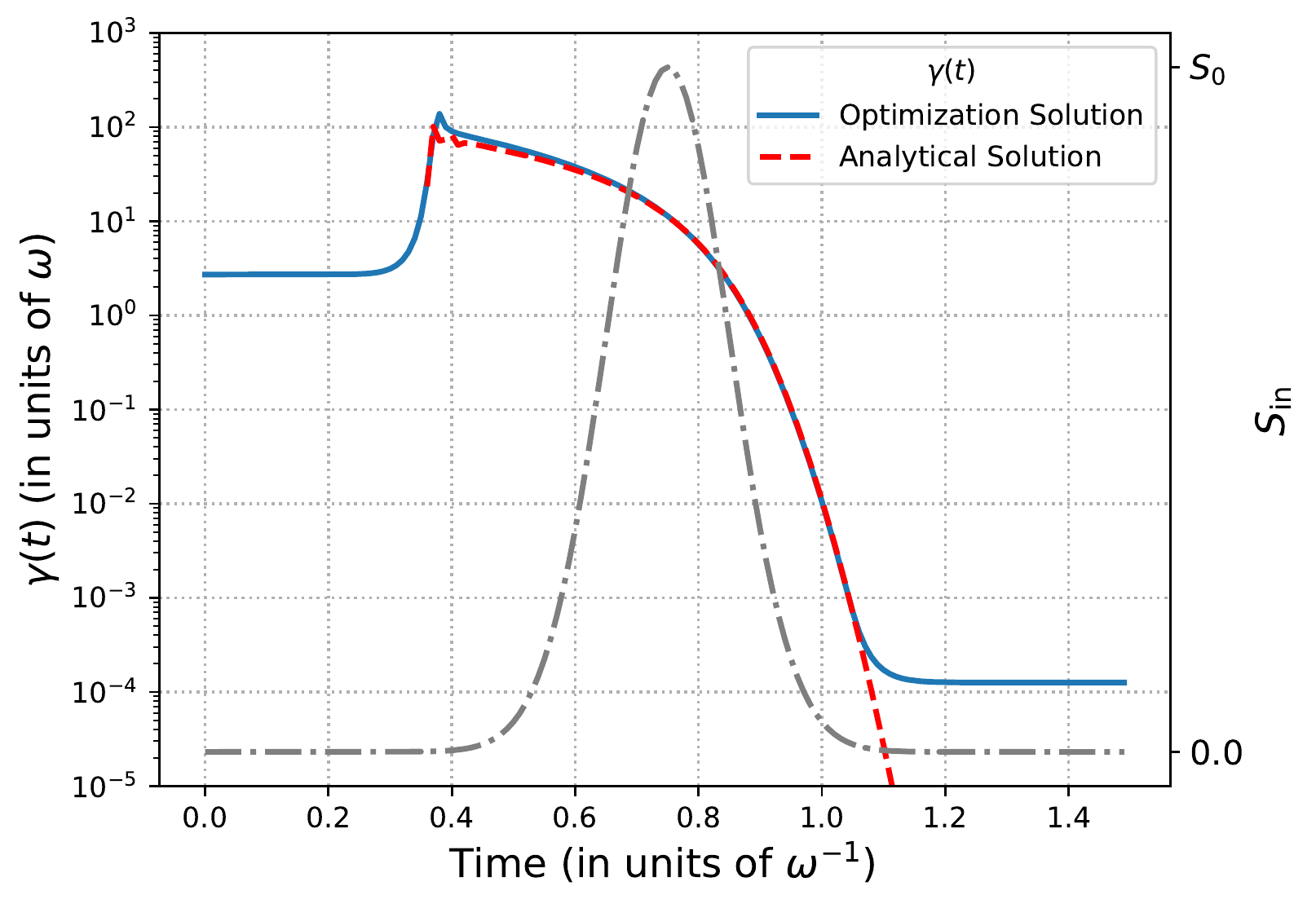}
\caption{\label{gamma_opt}The controllable coupling coefficient $\gamma(t)$ for the capture of an incoming pulse of a Gaussian envelope (in dashed grey line, on the right vertical axis).}
\end{figure}

For other envelopes, a numerical solution, either through solving the differential equation, or through an optimization problem minimizing $S_\mathrm{out}$ is also possible. Fig.~\ref{gamma_opt} illustrates the agreement between the analytical solution for $\gamma(t)$ in Eq.~\eqref{analytic_sol} and that obtained via a generic numerical optimization.

\subsection{Nonlinear Activation Function with Second Harmonic Pump used for Amplification}



The proposed nonlinear activation function in the main text uses a subharmonic mode that interacts with the neuron modes. This subharmonic mode operates at frequencies that are at half of the frequencies of the neuron modes. However, constraints such as the transparency of the material and the availability of high efficiency sources at required frequencies could present experimental difficulties. Hence, we discuss an alternative nonlinear activation function, where we allow the neuron modes to interact with pumps that are at the second harmonic. Similarly to the main text, the interaction can be modelled by the same system of partial differential equations, in which we permute the neuron and pump modes:
\begin{equation}
    \frac{\partial E_{\mathrm{sec}}}{\partial z} + \frac{\eta}{c} \frac{\partial E_{\mathrm{sec}}}{\partial t} = -\kappa E_{\mathrm{n}}^{2} - \alpha E_{\mathrm{sec}}
\end{equation}
\begin{equation}
    \frac{\partial E_{\mathrm{n}}}{\partial z} + \frac{\eta}{c} \frac{\partial E_{\mathrm{n}}}{\partial t} = \kappa E_{\mathrm{sec}} E_{\mathrm{n}}^{*} - \alpha E_{\mathrm{n}}
\end{equation}
where, $E_{\mathrm{n}}$ is the neuron mode and $E_{\mathrm{sec}}$ is the second harmonic pump mode.

These nonlinear activation functions can be used in order to amplify the neuron modes and circumvent losses.

\subsection{Equations of Motion in a Waveguide with Three-Wave-Mixing}

Here we explicitly derive the equations of motion used in the main text. Consider the "neuron" and "subharmonic" fields:
\begin{equation}
    \mathbf{E}_n = \mathbf{f}^p_n(x, y)E_n(z, t)e^{i(\omega t - kz)} + \mathrm{c.c.}
\end{equation}
\begin{equation}
    \mathbf{E}_s = \mathbf{f}^p_s(x, y)E_s(z, t)e^{i\frac{1}{2}(\omega t - kz)} + \mathrm{c.c.}
\end{equation}
Here $\mathbf{f}^p_*(x, y)$, describes the profile of the waveguide mode and $\mathbf{E}_*(z, t)$ describes the shape of the wave packet. Of note is that we keep track of the complex conjugate part as we have nonlinear processes. Given Maxwell's equations in matter and the typical parameterization of nonlinear susceptibility we have the nonlinear wave equation
\begin{equation}
    \left(\nabla^2-\frac{n^2}{c^2}\right)\mathbf{E} = \frac{1}{\varepsilon_0 c^2}\partial_t^2\mathbf{P}_\mathrm{NL} = \frac{1}{c^2}\chi^{(2)}\mathbf{E}\mathbf{E},
\end{equation}
where $\mathbf{E}=\mathbf{E}_n+\mathbf{E}_s$ and we have approximated $\nabla\cdot\textbf{E}=0$.

The linear version of the above equation provides an eigenvalue problem defining the shape of the waveguide modes:
\begin{equation}
\left(\partial_x^2+\partial_y^2\right)\mathbf{f}^p_n(x, y)=\left(-(ik)^2+(i\omega)^2\frac{n^2}{c^2}\right)\mathbf{f}^p_n(x, y)
\end{equation}
\begin{equation}
\left(\partial_x^2+\partial_y^2\right)\mathbf{f}^p_s(x, y)=\left(-\left(i\frac{k}{2}\right)^2+\left(i\frac{\omega}{2}\right)^2\frac{n^2}{c^2}\right)\mathbf{f}^p_s(x, y).
\end{equation}

The nonlinear perturbation leads to the following equation of motions for the wave packet envelopes. In its derivation we take into account that they are slowly varying functions for which $\partial_z\ll k$ and $\partial_t\ll \omega$.
\begin{equation}
\left(\partial_z+\frac{n}{c}\partial_t\right)E_n(z,t) = -\kappa E_s^2(z,t),
\end{equation}
\begin{equation}
\left(\partial_z+\frac{n}{c}\partial_t\right)E_s(z,t) = \kappa E_n(z,t)E_s^*(z,t),
\end{equation}
where $\kappa$ is
\begin{equation}
\kappa \frac{c}{\omega}
=\frac{\int \chi^{(2)}\mathbf{f}^{p*}_n\mathbf{f}^p_s\mathbf{f}^p_s \mathrm{d}x \mathrm{d}y}{
\int \mathbf{f}^{p*}_n\mathbf{f}^p_n \mathrm{d}x \mathrm{d}y}
=-
\frac{\int \chi^{(2)}\mathbf{f}^{p*}_s\mathbf{f}_n\mathbf{f}^{p*}_s \mathrm{d}x \mathrm{d}y}{
\int \mathbf{f}^{p*}_s\mathbf{f}^p_s \mathrm{d}x \mathrm{d}y}.
\end{equation}
These two expressions for $\kappa$ have to be equal for energy to be conserved in the equations of motion. For mode overlap on the order of unity we have $\kappa \frac{c}{\omega}\approx\chi^{(2)}$.

\end{document}